\documentclass[12pt]{article}
\usepackage{setspace}
\usepackage{amsmath}
\usepackage{graphicx}
\usepackage{cases}
\usepackage{subfig}
\usepackage{natbib}
\usepackage{booktabs}
\usepackage{psfrag}
\usepackage{amssymb}
\usepackage{color,graphics,epsfig}
\usepackage{appendix}
\usepackage{textcomp}
\usepackage{soul}
\usepackage{multirow}
\usepackage{xcolor}

\def\be{\begin{equation}}
\def\ee{\end{equation}}
\def\bea{\begin{eqnarray}}
\def\eea{\end{eqnarray}}
\def\bml{\begin{mathletters}}
\def\eml{\end{mathletters}}

\begin{document}

\title{Role of epistasis on the fixation probability of a {non-mutator} in an adapted asexual population}

\author{Ananthu James \\Theoretical
Sciences Unit, \\Jawaharlal Nehru Centre for Advanced Scientific
Research, \\ Jakkur P.O., Bangalore 560064, India} 

\maketitle

\newpage


\bigskip

\noindent
Corresponding author: \\
Ananthu James, \\
Theoretical Sciences Unit,\\
Jawaharlal Nehru Centre for Advanced Scientific Research, \\ 
Jakkur P.O., Bangalore 560064, India. \\
\texttt{ananthujms@jncasr.ac.in}

\bigskip
   
\noindent
\textbf{Abstract: The mutation rate of a well adapted population is prone to reduction so as to have {a} lower mutational load. {We aim to} understand the role of epistatic interactions between {the} fitness affecting mutations in this process. Using a multitype branching process, the fixation probability of a single {non-mutator} emerging in a large asexual mutator population is analytically calculated here. The mutator population undergoes deleterious mutations at constant, but {at a} much higher rate than that of the {non-mutator}. We find that antagonistic epistasis lowers the chances of mutation rate reduction, while synergistic epistasis enhances it. Below a critical value of epistasis, the fixation probability behaves {non-m}onotonically with variation in mutation rate of the background population. {Moreover}, the variation of this critical value of {the} epistasis parameter with the strength of the mutator is discussed {in the Appendix}. For synergistic epistasis, when selection is varied, {the fixation probability reduces overall, with damped oscillations.}} 


\newpage

\section{{Introduction}}\label{intro}

{Genetic variations in a population are essential for natural selection to act, resulting in increase in the number} of individuals more suited to the environment. This process is called adaptation \citep{Charlesworth:2010}. Mutation is one of the main sources of variation \citep{Charlesworth:2010}. {The mutation rate is defined to be the number of mutations occurring per cell division or per generation} \citep{Baer:2007}. Mutation rates being different for individuals of the same species and {amongst different species} \citep{Baer:2007} points to the fact that mutation rates are subject to the action of other evolutionary forces \citep{Raynes:2014}. Laboratory experiments reveal that owing to {the} ability to quickly generate beneficial mutations, and hitchhike with them, {higher mutation rate or mutator alleles are positively selected} in populations adapting to a new environment \citep{Smith:1974,Sniegowski:1997,Raynes:2014}. {Various theoretical studies have addressed hitchhiking in adapting populations} \citep{Taddei:1997,Tenaillon:1999,Johnson:1999a,Palmer:2006,Wylie:2009,Sniegowski:2010,Desai:2011}. 

{An experiment} by \citet{Giraud:2001} sheds light on the fact that mutators are no longer beneficial after adaptation. In fact, lower mutation rate or {non-mutator allele is favored} in populations that are adapted to an environment \citep{Trobner:1984,Notleymcrobb:2002,Mcdonald:2012,Turrientes:2013,Wielgoss:2013}, in order to have reduced load of deleterious mutations \citep{Liberman:1986}. Since beneficial mutations are found to be much rarer compared to deleterious mutations \citep{Drake:1998}, and adapted populations are assumed to be near their fittest genotype so as not to have space for further improvement, most of the theoretical studies on adapted populations have neglected the effect of beneficial mutations \citep{Lynch:2011,Soderberg:2011,Jain:2012}. However, \citet{James:2016} studied an asexual population at mutation{-}selection balance in which compensatory mutations are allowed. 

When {the} selective effects are much stronger than mutation rates, individuals with {non-zero} number of mutations will get lost from the population. Using this assumption,  \citet{Lynch:2011} addressed the problem of lowering of mutation rate in an adapted population. This is effectively a one locus model. \citet{James:2016} extended {this} study by relaxing the strong selection assumption. {They} analytically calculated the fixation probability of a {non-mutator} arising in a background which has very high mutation rate, using a multitype branching process \citep{Johnson:2002}. The current study aims to have a better understanding of the process of mutation rate reduction by further extending the approach of \citet{James:2016} when epistatic interactions are present. Although epistasis can have an impact on the transitions between mutators and {non-mutator}s by controlling the sites responsible for {the} change in mutation rate \citep{Wielgoss:2013}, the present article intends to be solely on epistatic interactions among {the} fitness affecting mutations.

Except \citet{Jain:2012}, all the works listed here on adapted populations considered mutations to contribute independently to fitness, which is otherwise known as a {non-e}pistatic fitness landscape. An epistatic landscape is {a} more general description of the actual biological scenario, since intergenetic interactions cannot be ignored. There have been numerous experiments demonstrating the presence of epistasis \citep{Mukai:1969,Whitlock:2000,Maisnier-Patin:2005,Kryazhimskiy:2009,Chou:2011,Khan:2011,Plucain:2014}. The effect of epistasis on asexual populations have been explored theoretically as well \citep{Kondrashov:1993,Wiehe:1997,Campos:2004,Jain:2007b,Jain:2008,Jain:2010,Jain:2012,Fumagalli:2015}. While \citet{Campos:2004} studied the process of fixation of a mutant with a direct selective advantage in a population that is undergoing deleterious mutations at constant rate, \citet{Jain:2012} explored the fixation of mutators. The focus in this article is to understand the fixation of {non-mutators, for which, the probability of fixation of \textit{a single non-mutator} is studied.}

In the current study, we find that synergistic epistasis (two or more mutations interact with each other to produce larger decline in relative fitness) rises {the} fixation probability of a rare {non-mutator}, whereas antagonistic or diminishing epistasis (two or more mutations interact with each other to produce smaller decline in relative fitness) lowers it. When selection is much stronger compared to mutation rate, {the} fixation probability is independent of epistasis, and increases with mutation rate. This matches with the result of \citet{James:2016} in the absence of epistasis. Below a particular value of antagonistic epistasis, we see that the fixation probability initially increases, and then decreases with mutation rate of the background. {In the presence of synergistic} interactions, as selection is varied, {the fixation probability decreases overall, with damped oscillations.} Our results can be merged with that of \citet{Kondrashov:1993} to deduce that synergistic epistasis is {doubly advantageous }as it {not only lowers} the rate of accumulation of deleterious mutations, but also increases the chances of mutation rate reduction. On the other hand, antagonistic epistasis is {doubly disadvantageous} to an asexual population due to the faster rate of accumulation of harmful mutations \citep{Wiehe:1997}, as well as {the} lower probability of mutation rate decline. 

\section{{Models and methods}}

\subsection{{Details of stochastic simulations}}\label{sto_sim}

We consider a large asexual population of haploid individuals of size $N$ on a fitness landscape \citep{Wiehe:1997}
\be 
\label{fitfunction}
W(k)=(1-s)^{k^{\alpha}} ~,~ 
\ee 
where $0$ $<$ $s$ $<$ $1$ is the selection coefficient and $\alpha$ $>$ $0$ is the epistasis parameter. Here, $k$ is the number of deleterious mutations carried by the genome, represented using a binary sequence of length $L$ $\rightarrow$ $\infty$, of an individual. We also denote $k$ as the fitness class{ since} the fitness is decided by $k$. Antagonistic epistasis is modeled by $\alpha$ $<$ $1$ and synergistic epistasis by $\alpha$ $>$ $1$. $\alpha$ $=$ $1$ implies no epistasis. Biologically, (\ref{fitfunction}) represents a genome carrying infinite number of biallelic loci that are equivalent to each other, and the effect of a new mutation at any locus depends on the number of mutations already present in the genome. The probability that the genome of an individual accumulates $x$ number of deleterious mutations at the rate $U_d$ is Poisson distributed as given below. 
\be
\label{mutprob}
M_{U_d}(k\rightarrow k+x) = e^{-U_d} ~ \frac{\left(U_d\right)^{x}}{x!} ~.~
\ee

The population evolves via standard Wright-Fisher (W-F) dynamics \citep{Jain:2008}, where the population size is held constant in each {non-o}verlapping generation. In {the} W-F process, corresponding to each individual, we randomly assign an individual in the previous generation as its parent{. This undergoes mutation} followed by reproduction with a probability proportional to its fitness. 

Asexual populations can go extinct via the accumulation of deleterious mutations (see section \ref{summary}), a process known as Muller's ratchet \citep{Haigh:1978,Kondrashov:1993}. Populations of large size with extremely small ratchet speed, that have been evolving for long timescales without changes in the environment can attain {a} steady state due to mutation-selection balance. For a population in steady state, the mean fitness and the population fractions corresponding to various genotypes remain time independent. In this study, it is assumed that the {non-mutator} with mutation rate $U'_d$ $=$ $U_d/\lambda$, where $\lambda$ $>$ $1$ is the strength of the mutator, appears when the mutator population is in steady state. The {non-mutator} also evolves via standard Wright-Fisher process, and (\ref{fitfunction}) and (\ref{mutprob}) are applicable for it with $U_d$ being replaced by $U'_d$. Here, we choose populations of size large enough to fulfill the criterion that {the number of individuals carrying the minimum number of mutations (least loaded class) in steady state} is at least $100${ so that} Muller's ratchet operates at a very small speed \citep{Kondrashov:1993} (also, see {Appendix} \ref{steadystate}).

In simulations, we consider the population to be {in} steady state initially. This assumption is verified by ensuring that the population eventually reaches mutation-selection balance by observing single run plots corresponding to the given parameter set of $s$, {$U_d$, and} $\alpha$. {Moreover}, we confirm that the population fractions stabilize at values predicted by (\ref{mutfreq_soln}). Fig. \ref{fpNsm} {in the Appendix} shows qualitative comparison of the fixation probability of a {non-mutator} created at time $t=0$ in a population that is in steady state ({filled} symbols) with that of a {non-mutator} produced after a time interval of $10/s$ generations in a population which initially has no deleterious mutations ({open} symbols). In this article, each simulation point (excluding the points in the single run plot Fig. \ref{syn_cl}) is averaged over $10^5$ independent stochastic runs. All simulations except those for Fig. \ref{fpNsm} have assumed $N$ $=$ $4,000$. Apart from Fig. \ref{crit_alpha}, only {non-mutator}s with $\lambda$ $=$ $100$ have been considered here. All the numerical calculations have been done using \textit{{Wolfram Mathematica}} $9.0.1.0$.

\subsection{{Analysis}}\label{analys}

Due to {the} lower rates of deleterious mutation accumulation and fitness decline, the {non-mutator} appearing in mutator background in an adapted population is effectively a beneficial allele. The fixation probability of such an allele can be studied using {the} branching process \citep{Patwa:2008}. The details \citep{Johnson:2002} are described below.

{The extinction} probability $\epsilon(k,t)$ of a {non-mutator} arising with $k$ deleterious mutations in generation $t$ in a very large population of mutators is given by 
\begin{eqnarray}
\label{extprob1} 
\epsilon(k,t) = \sum_{n=0}^{\infty} \psi_n (k,t) \left[\sum_{j} M_{U'_d}(k\rightarrow j) ~ \epsilon(j,t+1)\right]^n  ~.~
\end{eqnarray}
The above equation assumes that the extinction probabilities are independent of each other. Here, $\psi_n (k,t)$ is the probability that the {non-mutator} will give rise to $n$ offspring in generation $t$. $M_{U'_d}(k\rightarrow j)$ is the Poisson distributed probability of the {non-mutator} to mutate from class $k$ to $j$ $>$ $k$.

If the probability of reproduction of the {non-mutator} is assumed to be Poisson distributed, we get 
\be
\label{reprprob}
\psi_n (k,t) = e^{-w(k,t)} ~ \frac{w^n (k,t)}{n!} ~.~
\ee
In this expression, the mean of the Poisson distribution equals the absolute fitness of the {non-mutator}, and hence we write 
\be
\label{abs_fit}
w(k,t) = \frac{W(k)}{\overline W(t)} ~.~
\ee
Note that $\overline W(t)$ $=$ $\sum_{k=0}^{\infty} W(k) ~ p(k,t)$ is the mean fitness of the background population with $p(k,t)$ being the fraction of population having $k$ deleterious mutations in generation $t$ (see Appendix \ref{mutfreq} for details on the expression $p(k)$ for population fraction in steady state). With the help of (\ref{reprprob}) and (\ref{abs_fit}), we rewrite (\ref{extprob1}) as 
\begin{eqnarray}
\label{extprob2} 
\epsilon(k,t) =  e^{- \frac{W(k)}{\overline W(t)} \left[1 - \sum_{j} M_{U'_d}(k\rightarrow j) ~ \epsilon(j,t+1) \right]}  ~.~
\end{eqnarray}

The {non-mutator}s are considered established if they do not go extinct. Due to the selective advantage possessed by the {non-mutator, the establishment} eventually leads to fixation, and these two are taken to be the same here. {Hence, the fixation probability is} $\pi(k,t)$ $=$ $1-\epsilon(k,t)$. {Therefore, from} (\ref{extprob2}), {it follows that} 
\be
\label{fpit}
1 - \pi(k,t) = e^{- \frac{W(k)}{\overline W(t)} \sum_{i} e^{-U'_d} ~ \frac{(U'_d)^{i}}{i!} ~ \pi(i+k, t+1) }  ~,~
\ee 
{since} $\sum_{i=0}^{\infty} M_{U'_d}(k\rightarrow i+k)$ $=$ $1$. For a {non-mutator} that arises in the background population after the attainment of steady state, (\ref{fpit}) becomes 
\be
\label{fpk_rec}
1 - \pi(k) = e^{- \frac{W(k)}{\overline W} \sum_{i} e^{-U'_d} ~ \frac{(U'_d)^{i}}{i!} ~ \pi(i+k) }  ~.~
\ee 

We get the fixation probability of a {non-mutator} that is produced in a genetic background having $k$ number of deleterious mutations by solving (\ref{fpk_rec}). {However}, the mutator population is distributed across so many fitness classes, and the {non-mutator} can appear in any of these backgrounds. {Hence}, the total fixation probability can be calculated only {by taking into account} all the possible genetic backgrounds. The probability of the {non-mutator} to appear in fitness class $k$ is {the} same as the fraction $p(k)$ of {the} background population in that class. This is an important concept which plays a major role in understanding the results. As explained above, the \textit{total fixation probability} receives contributions from both {the \textit{fraction of background population}} and {the} probability of \textit{fixation}, and therefore, can be expressed as 
\be
\label{fptot}
\Pi = \sum_{k} p(k)~\pi(k) ~.~
\ee
The above expression is applicable for very large populations in which the effect of genetic drift can be neglected. 

\section{Results}\label{results}

As considered by \citet{James:2016}, for strong mutators which have very high mutation rates ($\lambda$ $\gg$ $1$) compared to the {non-mutator} \citep{Sniegowski:1997,Oliver:2000}, we can neglect $U'_d$ to write 
\be
\label{fpk_strongmut}
1 - \pi(k) = \exp\left[-\frac{W(k)}{\overline W} \pi(k)\right]~.~
\ee  
If the {non-mutator} has negligible mutation rate, we can directly obtain (\ref{fpk_strongmut}) from (\ref{extprob1}) assuming steady state, since $\pi(k)$ $=$ $1-\epsilon(k)$. Using (\ref{meanmutexp}), the average fitness of mutators in steady state is found to be $\overline W = (1-s)^{\overline{k^{\alpha}}} \approx e^{-U_d}$. This is otherwise the classical result obtained by \citet{Haldane:1937} for the mean fitness of an asexual population. Following the approach of \citet{James:2016}, taking logarithm on both sides of (\ref{fpk_strongmut}), and neglecting terms of order greater than $2$ from the expansion $\ln (1-x) = - x - x^2/2 - ...${,} we can solve the resulting quadratic equation to get
\begin{equation}
\pi(k) =
\begin{cases}
2\left(\frac{W(k)}{\overline W}-1\right) ~~ = ~~ 2s\left(\overline{k^{\alpha}}-k^{\alpha}\right) ~~ \text{if} \hspace{0.05 cm} ~~ k<\lfloor (U_d/s)^{1/\alpha} \rfloor \\
0 ~~ \text{otherwise} \hspace{0.05 cm} ~.~ 
\end{cases}
\label{fpk}
\end{equation}
Here, $\lfloor (U_d/s)^{1/\alpha} \rfloor$ is the largest integer corresponding to $(U_d/s)^{1/\alpha}$, and $\overline{k^{\alpha}}$ is given by (\ref{meanmutexp}). We see that with rise in {the} background mutation rate $U_d$, $\pi(k)$ increases, which is rather expected. The intuitive meaning of (\ref{fpk}) is that the effective selective advantage of a {non-mutator} carrying $k$ mutations, appearing in the background having mean fitness $e^{-s\overline{k^{\alpha}}}$ is $s(\overline{k^{\alpha}}-k^{\alpha})$, and its fixation probability is twice that. The latter statement follows from {the} single locus model \citep{Fisher:1922,Haldane:1927}. 

Plugging (\ref{mutfreq_soln}) and (\ref{fpk}) in (\ref{fptot}), and performing the resulting sum give rise to 
\be
\label{fptot_full}
\Pi = \frac{2U_d ~ (U_d/s)^{\lfloor (U_d/s)^{1/\alpha} \rfloor}}{\left({\lfloor (U_d/s)^{1/\alpha} \rfloor}!\right)^\alpha} ~ p(0) ~.~
\ee
{The derivations} for the frequency $p(0)$ of the background population with zero deleterious {mutation} are given in {Appendices} \ref{mutfreq} {and} \ref{mutfreq0}, and the final expressions are summarized in Table \ref{tablep0}. Based on whether the selection is strong ($U_d/s < 1$) or weak ($U_d/s > 1$){ and} the epistasis is antagonistic ($\alpha < 1$) or synergistic ($\alpha > 1$), there are {four} regimes for $\Pi$.

\begin{table}
\begin{center}
  \begin{tabular}{| l || l | l | }
\hline
\multicolumn{3}{ |c| }{\bf{Class $0$ mutator frequency $p(0)$}} \\
    \hline
    ~ & $\left(\frac{U_d}{s}\right)$ $>$ $1$  \vspace{0.0 cm} & $\left(\frac{U_d}{s}\right)$ $<$ $1$ \\ \hline \hline
    $\alpha$ $\leq$ $1$ & $p(0)=(2 \pi)^{\frac{\alpha-1}{2}} \frac{ \sqrt{\alpha}\left(\frac{U_d}{s}\right)^{\frac{\alpha - 1}{2 \alpha}}}{ e^{\alpha \left(\frac{U_d}{s}\right)^{1/\alpha}}} $  \vspace{0.0 cm} & $p(0) = \left(1-\frac{U_d}{s}\right) ~~ \text{if \hspace{0.05 cm}} ~~ \alpha \ll 1$ \vspace{0.0 cm} \\ \hline

    $\alpha$ $>$ $1$ & ${p(0) = \left(1+\frac{U_d}{s}+\frac{(U_d/s)^2}{2^{\alpha}}\right)^{-1}} $ & ${p(0) = \left(1+\frac{U_d}{s}+\frac{(U_d/s)^2}{2^{\alpha}}\right)^{-1}} $ \\
     & $~ \text{if \hspace{0.05 cm}} ~~  \alpha > \frac{\ln{(U_d/s)}}{\ln{2}}$ & \\
    \hline

    {$\alpha$ $=$ $2$} & ${p(0) = \left[I_0\left(2\sqrt{\frac{U_d}{s}}\right)\right]^{-1}}$ & ${p(0) = \left[I_0\left(2\sqrt{\frac{U_d}{s}}\right)\right]^{-1}}$ \\ \hline
  \end{tabular}
\caption {The above expressions are derived in {Appendices} \ref{mutfreq} {and} \ref{mutfreq0}. Expressions in the last row are exact, while the other ones are approximations. The symbols $U_d$, $s$ and $\alpha$ respectively represent {the} mutation rate of the background population, selection coefficient and epistasis parameter.}    
\label{tablep0}
\end{center}
\end{table} 

\subsection{Variation of fixation probability with epistasis parameter}\label{res_ep}

\subsubsection{{Weak selection; antagonistic epistasis ($U_d/s$ $>$ $1$, $\alpha$ $\leq$ $1$)}}\label{wsae}

For large $(U_d/s)^{1/\alpha}$, with the help of Stirling's approximation $x! \approx \sqrt{2\pi x} ~ (x/e)^x$, we obtain 
\be
\label{fptot_exp}
\Pi = \frac{2U_d \left( U_d/s\right)^{(1-\alpha)U_d/s} e^{U_d\alpha/s}}{\left(2\pi U_d/s\right)^{\alpha/2}} ~ p(0) ~.~
\ee
Using the result from Table \ref{tablep0}, we get  
\be
\label{fptot_exp_ws_al1}
\Pi = U_d ~ \sqrt{\frac{2\alpha}{\pi}} ~ \left(\frac{s}{U_d}\right)^{\frac{1}{2\alpha}} ~.~
\ee
Expression (\ref{fptot_exp_ws_al1}) yields the known result \citep{James:2016} for $\alpha=1$. It is evident that $\Pi$ $\propto$ $U_d^{1-\frac{1}{2\alpha}}$, implying the total fixation probability to be an increasing function of the background mutation rate for $\alpha$ $>$ $0.5$ and decreasing function for $\alpha$ $<$ $0.5$, as shown in Fig. \ref{fpsm}. The value of $\alpha$ at which this transition happens is denoted as $\alpha_c$, the critical value of {the} epistasis parameter. Corresponding to $\alpha$ $=$ $\alpha_c$, (\ref{fptot_exp_ws_al1}) gives $\Pi$ $=$ $\frac{s}{\sqrt{\pi}}$. As the mutation rate of a population increases, we expect it to have higher probability to reduce its mutation rate. {However, if $\alpha$ $<$ $0.5$, we see that the higher the mutation rate of a population, the lower is the probability that its mutation rate will decrease.} The physical interpretation of this surprising trend is explained in the following paragraph. 

Combining (\ref{p0_gaussian}) and (\ref{mutfreq_0_simple}) enables us {to} write 
\be
\label{pkgaussain}
p(k) = \frac{ e^{\frac{-\alpha \left(k-(U_d/s)^{1/\alpha}\right)^2}{2(U_d/s)^{1/\alpha}}} }{\sqrt{2 \pi ~ \frac{(U_d/s)^{1/\alpha}}{\alpha} }} ~.~
\ee
This clearly states that the background population frequency $p(k)$, which is also equal to the probability of the {non-mutator} to appear with $k$ deleterious mutations, is a Gaussian distribution with mean $(U_d/s)^{1/\alpha}$ and variance ${\alpha}^{-1} (U_d/s)^{1/\alpha}$. Therefore, in the regime $\alpha$ $<$ $1$ and $(U_d/s)$ $>$ $1$, the mutator population will be more spread out for larger values of $U_d$ and smaller values of $\alpha$. 

As $U_d$ increases, it is more likely that the {non-mutator} will appear with higher number of deleterious mutations, which is disadvantageous to the invader population. {However}, as we saw in (\ref{fpk}), once the {non-mutator} appears with a particular number of mutations, its fixation probability $\pi(k)$ increases with $U_d$. {This} is an advantageous factor associated with $U_d$. 
Competition between the advantageous and disadvantageous effects of $U_d$ on the {non-mutator} decides the behavior of its total fixation probability as a function of $\alpha$. As $\alpha$ falls below $0.5$, the disadvantage experienced by the lower mutation rate allele due to its low fitness dominates its advantage of arising in a background that has high mutation rate. 

Fig. \ref{fpsm}, \ref{fpnonmon_al1}{, and} \ref{fps__al1} show that the trend predicted by (\ref{fptot_exp_ws_al1}) is observed in finite size populations. {Further}, in the parameter regime used in the plot, (\ref{fptot_exp_ws_al1}) is a good approximation for the total fixation probability of a lower mutation rate individual in {the} strong mutator background. 

{It has to be noted that the result $\alpha_c$ $=$ $0.5$ is valid only for the strong mutator background. A discussion on $\alpha_c$ for the case in which the mutation rates of the non-mutator and mutator are comparable (weak mutator background) can be found in Appendix} \ref{alphaC_wm}.   

\subsubsection{{Weak selection; synergistic epistasis ($U_d/s$ $>$ $1$, $\alpha$ $>$ $1$)}}\label{wsse}

We have {an} analytical expression for $p(0)$ only for the limiting case $\alpha$ $>$ $\frac{\ln{(U_d/s)}}{\ln{2}}$ ($(U_d/s)^{1/\alpha}$ $<$ $2$), which is given in Table \ref{tablep0}. {The mutation rates} of asexual microbes such as \textit{{E. coli}} and \textit{{S. cerevisiae}} are measured to be of the order of $10^{-3}$ per genome per generation \citep{Drake:1998}. The value of selection coefficient for \textit{{E. coli}} is found to vary from $10^{-3}$ \citep{Gallet:2012} to $10^{-1}$ \citep{Lenski:1991}. Even for the maximum value of $(U_d/s)$ in this case, which is 100, $\alpha$ $\geq$ $6.7$ ensures that only the first two classes contribute to $\Pi$. In fact, even if $(U_d/s)$ $\sim$ $10^6$, which could be biologically improbable, $\alpha$ $\geq$ $20$ guarantees that $\pi(k)$ $=$ $0$ for $k$ $>$ $1$. This physically corresponds to {two or more mutations} interacting with each other {to produce} lethal effects on the genome. This is illustrated in Fig. \ref{syn_cl} {in the Appendix}. Therefore, it follows from (\ref{fptot_full}) that 
\be
\label{fptot_ws_ag1}
{\Pi = \frac{2 U_d ~ U_d/s}{\left(1+\frac{U_d}{s}+\frac{(U_d/s)^2}{2^{\alpha}}\right)} ~~ \text{if \hspace{0.05 cm}} ~~ \alpha > \frac{\ln{(U_d/s)}}{\ln{2}}}  ~.~
\ee
Here, the $\alpha$ dependence of $\Pi$ comes from $p(0)$. {For $\alpha$ $\gg$} $\frac{\ln{(U_d/s)}}{\ln{2}}$, {epistasis affects neither the fixation probability $\pi(k)$ nor the fraction of background population $p(k)$ of the first two fitness classes. Unsurprisingly, it can be seen that $\Pi$ rises with increase in $\alpha$ initially and reaches its maximum value} $2 U_d ~ (U_d/s) \left(1+\frac{U_d}{s}\right)^{-1}$, which is independent of $\alpha$. This is captured in Fig. \ref{fpsm} ({also, see Appendix} \ref{limit_ana_res} where the discrepancy between {the results from simulations and analytics has been discussed). It is important to note that} if the condition $\alpha$ $>$ $\frac{\ln{(U_d/s)}}{\ln{2}}$ is not satisfied, the {non-mutator} can arise in fitness classes having low fitness. Owing to this, (\ref{fptot_ws_ag1}) overestimates the actual value of $\Pi$. 

Note that for $(U_d/s)$ $\gg$ $1$ and $\alpha$ $\gg$ $\frac{\ln{(U_d/s)}}{\ln{2}}$, (\ref{fptot_ws_ag1}) simplifies to the known result for fixation probability $\Pi$ $=$ $2 U_d$ \citep{James:2016} of a {non-mutator} on a {non-e}pistatic fitness landscape, when {the} selective effects are strong with respect to mutation rate. From (\ref{mutfreq_soln}) and (\ref{mutfreq_0_ag1}), we see that synergistic epistasis with $\alpha$ $\gg$ $\frac{\ln{(U_d/s)}}{\ln{2}}$ causes the background population to be concentrated around fitness class $1$, and therefore, $p(1)$ $\approx$ $1$ for $(U_d/s)$ $\gg$ $1$. The fixation probability of a {non-mutator} with a single deleterious mutation is $\pi(1)$ $\approx$ $2U_d$ from (\ref{fpk}) when $(U_d/s)$ $\gg$ $1$. Thus, both $p(1)$ and $\pi(1)$ give the same results as $p(0)$ and $\pi(0)${, respectively} when selection is very strong and epistasis is either absent \citep{James:2016} or synergistic (see section \ref{ssse}). Effectively, {the fitness} class $1$ for $\alpha$ $\gg$ $\frac{\ln{(U_d/s)}}{\ln{2}}$ and $(U_d/s)$ $\gg$ $1$ ``replaces'' {fitness} class $0$ for $(U_d/s)$ $\ll$ $1$ and $\alpha$ $\geq$ $1$. {Fig.} \ref{fps__ag1} shows variation of (\ref{fptot_ws_ag1}) with $s$.  

\subsubsection{{Strong selection; antagonistic epistasis ($U_d/s$ $<$ $1$, $\alpha$ $<$ $1$)}}\label{ssae}

As $(U_d/s)$ $<$ $1$, $\lfloor (U_d/s)^{1/\alpha} \rfloor$ $=$ $0$, and hence {the} fixation probability receives contribution only from class $0$. Using the result from Table \ref{tablep0} in (\ref{fptot_full}), we obtain
\be
\label{fptot_ss_al1}
\Pi = 2 U_d ~ (1-U_d/s) ~~ \text{if \hspace{0.05 cm}} ~~ \alpha \ll 1  ~.~
\ee
Fig. \ref{fpsm_ss}, \ref{fpnonmon_al1} {and} \ref{fps__al1} show the validity of (\ref{fptot_ss_al1}) {by comparing against finite population simulations. For $(U_d/s)$ values comparable to $1$, if the condition $\alpha$ $\ll$ $1$ is not fulfilled, the expression for $p(0)$ is not valid (see Table} \ref{table_p0_ssal1}, {and Case III in Appendix} \ref{mutfreq0}). {From Fig.} \ref{fpnonmon_al1} {and} \ref{fps__al1}, we can infer that for $(U_d/s)$ $\gtrsim$ $0.5$, $\Pi$ varies with $\alpha$. For $(U_d/s)$ $\ll$ $1$, $\Pi$ is independent of $\alpha$. 

\subsubsection{{Strong selection; synergistic epistasis ($U_d/s$ $<$ $1$, $\alpha$ $>$ $1$)}}\label{ssse}

Since $\lfloor (U_d/s)^{1/\alpha} \rfloor$ $=$ $0$, the use of the result from Table \ref{tablep0} in (\ref{fptot_full}) yields
\be
\label{fptot_ss_ag1}
{\Pi = \frac{2 U_d}{\left(1+\frac{U_d}{s}+\frac{(U_d/s)^2}{2^{\alpha}}\right)}}  ~.~
\ee
{As in the case of} (\ref{fptot_ws_ag1}), with increase in $\alpha$, the dependence of $\Pi$ on epistasis will vanish, and (\ref{fptot_ss_ag1}) will approach the constant value $2 U_d\left(1+\frac{U_d}{s}\right)^{-1}$. Fig. \ref{fpsm_ss} and \ref{fps__ag1} show the comparison of (\ref{fptot_ss_ag1}) with finite population simulations (a discussion on the discrepancy between {the results from simulations and analytics} can be found in {Appendix} \ref{limit_ana_res}). Table \ref{tablePi} {gives summary of all the results from section} \ref{results}.

It is obvious that, for $(U_d/s)$ $\ll$ $1$, (\ref{fptot_ss_al1}) and (\ref{fptot_ss_ag1}) approach the known result for fixation probability $\Pi$ $=$ $2 U_d$ \citep{James:2016} in the absence of epistasis ($\alpha$ $=$ $1$). This can also be obtained using {a} single locus model, since the population consists only of class $0$ individuals, and the selective advantage of {non-mutator}s is the difference in {the} class $0$ frequencies. {As the} class {$0$} individuals remain unaffected by epistasis for very strong selection, $\Pi$, which receives contribution only from class {$0$}, is independent of $\alpha$.  

\subsection{{Variation of fixation probability with mutation rate}}\label{res_mut}

\subsubsection{{Strong selection; antagonistic epistasis ($U_d/s$ $<$ $1$, $\alpha$ $<$ $1$)}}\label{ssae_mr}

When we vary $U_d$ keeping $s$ to be the same, for $(U_d/s)$ $<$ $1$, only {the} class $0$ individuals decide $\Pi$. For $(U_d/s)$ $\ll$ $1$, $p(0)$ $\approx$ $1$ (see Table \ref{tablep0}). This means that a mutation is very costly, due to which any individual carrying it will not survive. As the number of mutators in class $0$ decreases with {the} rise in $U_d$, a reduction in {the} mutation rate will be highly favored. Thus, $\Pi$ increases with $U_d$. For $(U_d/s)$ $\gtrsim$ $0.5$, $\Pi$ depends on epistasis (Fig. \ref{fpnonmon_al1} and \ref{fps__al1}), {though the dependence is not analytically captured. This is because $p(0)$ depends on $\alpha$. {An} increase (decrease) in $U_d$ ($\alpha$) leads to decrease in $p(0)$ (see Table} \ref{table_p0_ssal1} {and Fig.} \ref{Q}), as the background population spreads out more. On the other hand, {$\pi(0)$ $=$ $2U_d$.} This results in $\Pi$ $=$ $\pi(0)p(0)$ showing $\alpha$ dependent behavior for $1$ $>$ $(U_d/s)$ $\gtrsim$ $0.5$ similar to that in the regime $(U_d/s)$ $>$ $1$, $\alpha$ $<$ $1$. 

\subsubsection{{Weak selection; antagonistic epistasis ($U_d/s$ $>$ $1$, $\alpha$ $<$ $1$)}}\label{wsae_mr}

As discussed in section \ref{wsae}, {the total fixation probability in the regime $(U_d/s)$ $>$ $1$ is a multilocus problem. For antagonistic epistasis, the {non-mutator} has higher chances of appearing in a lower fit background for larger values of $U_d$. When $\alpha$ $<$ $0.5$, this disadvantage cannot be compensated by its benefit associated with being created in {a} higher mutation rate background. Due to this, $\Pi$ falls as a function of $U_d$. These two factors together give rise to a {non-m}onotonic behavior of $\Pi$ with respect to $U_d$ for $\alpha$ $<$ $0.5$, as shown in Fig.} \ref{fpnonmon_al1}. If $1$ $\geq$ $\alpha$ $>$ $0.5$, we see that the advantage conferred by the {non-mutator} owing to being produced in a high mutation rate background dominates its drawback and {therefore}, $\Pi$ increases with $U_d$. Thus, $\Pi$ is a monotonically increasing function of $U_d$ for $1$ $\geq$ $\alpha$ $>$ $0.5$ (see Fig. \ref{fpnonmon_al1}). 

\subsubsection{{Synergistic epistasis ($\alpha$ $>$ $1$)}}\label{se_mr}

For synergistic epistasis, $\Pi$ rises with $U_d$ for both weak selection (see Fig. \ref{fpsm}) and strong selection (see (\ref{fptot_ss_ag1})). {For $\alpha$ $\gg$} $\frac{\ln{(U_d/s)}}{\ln{2}}$, $\Pi$ is a linearly increasing function of $U_d$ for {both $(U_d/s)$ $\gg$ $1$ and $(U_d/s)$ $\ll$ $1$.}  

\subsubsection{{Effect of very large or small $(U_d/s)$ when $s$ is held constant}}

As $U_d$ increases to very large values relative to selection, for $\alpha$ $<$ ($>$) $\alpha_c$, $\Pi$ approaches $0$ ($1$). Note that $\Pi$ can never exceed $1$. When $U_d$ is decreased to very small values compared to $s$, $\Pi$ falls towards $0$ irrespective of $\alpha$. This is obvious because {the} lower the mutation rate of the mutator {is}, {the} smaller {the} advantage associated with {the reduction of mutation rate}. 

\subsection{{Variation of fixation probability with selection}}\label{res_sel}

\subsubsection{{Antagonistic epistasis ($\alpha$ $<$ $1$)}}\label{ae_sel}

The selection coefficient decides the effect of a mutation. Since the expression (\ref{fptot_ss_al1}) for $(U_d/s)$ $<$ $1$ and $\alpha$ $<$ $1$ is inadequate to capture the $\alpha$ dependence, {the} simulation data corresponding to two $\alpha$ values have been plotted for that regime in Fig. \ref{fps__al1}. Qualitatively, one can conclude that $\Pi$ increases with $s$. When $s$ is large, the {non-mutator} has a higher advantage by virtue of the higher deleterious effect of a mutation. For $(U_d/s)$ $\ll$ $1$, it is possible to understand from (\ref{fptot_ss_al1}) that $\Pi$ becomes independent of $s$, since it is determined only by {the} individuals that do not undergo mutation. In the weak selection regime, {better} understanding is possible with the help of (\ref{fptot_exp_ws_al1}), which is plotted in Fig. \ref{fps__al1} for $\alpha$ $=$ $0.6$. When $\alpha$ $=$ $0.2$, for the given set of parameters and population size ($N=4,000$), a steady state does not exist in the weak selection regime. {Hence}, the corresponding data is not shown in Fig. \ref{fps__al1}.

\subsubsection{{Synergistic epistasis ($\alpha$ $>$ $1$)}}\label{se_sel}

For synergistic epistasis, {in the special case of} $\alpha$ $=$ $2$, an exact solution exists for $p(0)$. With the help of the result in Table \ref{tablep0}, (\ref{fptot_full}) becomes
\be
{\Pi = \frac{2U_d ~ \left(\frac{U_d}{s}\right)^{\lfloor \sqrt{(U_d/s)} \rfloor}}{\left(\lfloor \sqrt{(U_d/s)} \rfloor !\right)^2 I_0\left(2\sqrt{\frac{U_d}{s}}\right)}} ~,~
\label{fp_a2}
\ee
which is valid for {any value} of $(U_d/s)$. Corresponding to {$(U_d/s)$ $<$ $1$}, $\lfloor \sqrt{(U_d/s)} \rfloor$ $=$ $0$, by which (\ref{fp_a2}) {takes the form} 
\be
{\Pi = \frac{2U_d }{ I_0\left(2\sqrt{\frac{U_d}{s}}\right)} ~~ \text{if \hspace{0.05 cm}} ~~ (U_d/s) < 1} ~.~
\label{fp_a2_ss}
\ee
Fig. \ref{fps__ag1} shows {the} comparison of (\ref{fp_a2}) with simulation data, {which is} represented using {the} blue open triangles. For {the} values of $s$ corresponding to which a population of size $N$ $=$ $4,000$ do not have steady state, the numerical solutions of (\ref{fpk_rec}) and (\ref{fptot}) using \textit{{Wolfram Mathematica}} $9.0.1.0$ are shown using {the} filled circles in the main figure {in} Fig. \ref{fps__ag1}. Since we assumed $U'_d$ $=$ $0${ and} used the approximation (\ref{fpk}), (\ref{fp_a2}) overestimates the exact numerical solution by a small amount. The comparison of (\ref{fp_a2}) with the exact solution using (\ref{fpk_rec}) and (\ref{fptot}) is given in the last {two} columns of Table \ref{table_synws}. {The detailed explanation} for the surprising trend in Fig. \ref{fps__ag1} is given in {Appendix} \ref{trend_fig6}.

\subsubsection{{Effect of very large or small $(U_d/s)$ when $U_d$ is kept the same}}

We know from sections \ref{ssae} and \ref{ssse} that when $(U_d/s)$ $\ll$ $1$, $\Pi$ assumes the value $2U_d$ regardless of $\alpha$. Expressions (\ref{fp_a2_ss}) and (\ref{fp_a1}) also justify this claim, as the denominators of them {approach $1$} for $(U_d/s)$ $\ll$ $1$. When selection is reduced to much lower values compared to mutation rate, $\Pi$ $\rightarrow$ $0${. This can be seen from} (\ref{fptot_exp_ws_al1}) for $\alpha$ $\leq$ $1${,} and Fig. \ref{fps__ag1} for $\alpha$ $=$ $2$. {This is because there will not be any significant difference between mutators and non-mutators when the selective effects are negligibly small.}

\begin{table}
\begin{center}
  \begin{tabular}{|| l || l | l ||}
\multicolumn{3}{ c }{\bf{{Analytical expressions for $\Pi$}}} \\
\multicolumn{3}{ c }{} \\
\toprule  

 ~ & ~ & ~ \\

    ~ &$(U_d/s)$ $\geq$ $1$ ~ ({Fig.} \ref{fpsm}) \vspace{0.0 cm} & $(U_d/s)$ $<$ $1$ ~ ({Fig.} \ref{fpsm_ss}) \\  \hline \hline
    
     $\alpha$ $\leq$ $1$ & $ \Pi = U_d ~ \sqrt{\frac{2\alpha}{\pi}} ~ \left(\frac{s}{U_d}\right)^{\frac{1}{2\alpha}} $  \vspace{0.0 cm} & $\Pi = 2 U_d ~ (1-U_d/s)$ \\ 

     ({Fig.} \ref{fpnonmon_al1} $\&$ \ref{fps__al1}) &  &  $\text{if \hspace{0.05 cm}} ~~ \alpha \ll 1$ \vspace{0.0 cm}\\ \hline

     $\alpha$ $>$ $1$ & ${\Pi = \frac{2 U_d ~ U_d/s}{\left(1+\frac{U_d}{s}+\frac{(U_d/s)^2}{2^{\alpha}}\right)}}$ & ${\Pi = \frac{2 U_d}{\left(1+\frac{U_d}{s}+\frac{(U_d/s)^2}{2^{\alpha}}\right)}} $ \\

     ({Fig.} \ref{fps__ag1}) &  $\text{if \hspace{0.05 cm}} ~~  \alpha > \frac{\ln{(U_d/s)}}{\ln{2}}$ & \\ \hline \hline

     {$\alpha$ $=$ $2$} & $ {\Pi = \frac{2U_d ~ \left(\frac{U_d}{s}\right)^{\lfloor \sqrt{(U_d/s)} \rfloor}}{\left(\lfloor \sqrt{(U_d/s)} \rfloor !\right)^2 I_0\left(2\sqrt{\frac{U_d}{s}}\right) }} $  \vspace{0.0 cm} &  $ {\Pi = \frac{2U_d }{ I_0\left(2\sqrt{\frac{U_d}{s}}\right)}} $ \vspace{0.0 cm} \\ 

     ({Fig.} \ref{fps__ag1}) &  & \\ 

\bottomrule  
  \end{tabular}
\caption {The figure denoted in bracket in any row (column) indicates that {the} validity of the next two expressions given in the same row (column) is shown in that figure. All expressions in the last column (strong selection regime) involve {non-mutator}s created only in class $0$. Even though the expressions in the second row can be used for $\alpha$ $=$ $2$, the equations in the last row are more accurate.}    \label{tablePi}
\end{center}
\end{table} 

\section{{Discussion}}\label{Discussion}

\subsection{{Summary of results and connection with real populations}}\label{summary}

In an asexual adapted population, it is known that \citep{Jain:2008} in {the} presence of synergistic epistasis, {a} higher proportion of individuals will carry less mutations, while in {the} presence of antagonistic epistasis, {the} fraction of individuals containing less mutations will be very low. In this article, it is found that antagonistic epistasis lowers the fixation probability of a lower mutation rate allele, thereby opposing {the decline in mutation rate}, while synergistic epistasis favors the reduction {in} mutation rate. Using a model similar to {the one} in this article, \citet{Kondrashov:1993} hypothesized that asexual populations could resist mutation accumulation in {the} presence of synergistic interactions. {However}, the model here assumes {the} population size to be a constant in every generation, while the size of an actual population fluctuates stochastically. A theoretical study in which the population size was allowed to vary with time \citep{Gabriel:1993} revealed that {the} extinction time of asexuals by virtue of onslaught of detrimental mutations vary depending on the values of parameters such as mutation rate, selection{,} etc. Since epistasis influences mutation rate reduction, results in the present article can be connected to the analysis of \citet{Gabriel:1993} to unravel the role of epistasis on {the} extinction time of asexuals. {However}, real populations can have compensatory or back mutations (see two different models that incorporate compensatory mutations - \citet{John:2015,James:2016}) acting against the influx of deleterious mutations, which were neglected by all the above studies. {In order to understand the fate of real asexual populations, the current study has to be extended by including more realistic considerations as discussed above.}    

It is observed in this study that there exists a critical value $\alpha_c$ of epistasis below which the probability of reduction of mutation rate in an infinite population shows negative correlation with its mutation rate. For strong mutators, $\alpha_c$ is found to be {at} $0.5$ using analytical arguments (see {Appendices} \ref{limit_ana_res} and \ref{alphaC_wm}). Though the mutation rate reduction happens with a {non-zero} probability, which is characteristic of any beneficial mutation, {the} decline in mutation rate becomes more unlikely when $\alpha$ $<$ $\alpha_c$. For $\alpha$ $<$ $\alpha_c$, {the} fixation probability of a {non-mutator} decreases with $U_d$ in the weak selection regime, as the lower mutation rate allele now appears with large number of mutations{ and} finds it difficult to outcompete the resident population. On the other hand, $\Pi$ increases with $U_d$ in the strong selection regime because of the {non-mutator} arising in class $0$ benefiting from {the} reduction of mutational load by {a} larger amount. {(A discussion on $\alpha_c$ in the weak mutator background is given in Appendix} \ref{alphaC_wm}.) 

In {the} presence of synergistic interactions, $\Pi$ initially remains constant at the value {$2U_d$} followed by a reduction, as $s$ is reduced in the strong selection regime. In {the} weak selection regime, as $s$ drops, $\Pi$ manifests a {non-m}onotonic behavior for every $n^{-\alpha}~U_d$ $<$ $s$ $<$ $(n+1)^{-\alpha}~U_d$, {where $n$ $=$ $1,2,3,...$, but experiences eventual decay. This is shown in Fig.} \ref{fps__ag1}.

\subsection{{Limitations of the models} and future goals}\label{limit_assum_model} 

One major assumption {which we have} made in this article is that the population attains steady state after a large number of generations. {However}, populations of small size (see {section} \ref{sto_sim} and {Appendix} \ref{steadystate}) do not have steady state, and hence the results presented here {are not applicable} to them. It is possible to evaluate the total fixation probabilities of {non-mutator}s in such populations using (\ref{fpit}) and (\ref{mutfreq_t}). The total fixation probability will depend on the time of arrival of the {non-mutator}. This has been studied by \citet{Lynch:2011} for a {non-e}pistatic landscape, when {the} selective effects are very strong. It is an open question for the epistatic landscape, and is not done here. The analysis in this article is a special case of this problem. 

{However, Fig.} \ref{fpNsm} {in the Appendix helps to have a qualitative understanding} of the effect of variation of $N$. For each value of $N$ in the figure, the {non-mutator} appears after $10/s$ generations. It is noticeable that large populations are more effective in withstanding the accumulation of harmful mutations. Small populations that are incapable of resisting the build-up of deleterious mutations and decline of mean fitness will be benefited more from {the reduction in mutation rate}. As a result, {when} $N$ increases, {the} fixation probability decreases{ and} approaches the constant value predicted by (\ref{fptot}). Articles like \citet{Lynch:2010b,Lynch:2011,Jain:2012,Sung:2012,James:2016}, etc. give insights {into} the role of population size on mutation rate evolution.

It is noteworthy that the model in this article examines the fixation probability of a single {non-mutator} appearing in the mutator population{ as} described using (\ref{extprob1}). {Its fixation is decided by whether the lineage of this particular non-mutator takes over the population or not}. In reality, there can be multiple {non-mutator}s emerging in the population. The article by \citet{Johnson:2002} includes the study of fixation probability of beneficial alleles arising at a constant rate. The same approach can be used to model the fate of multiple {non-mutator}s arising in the adapted mutator population, since the {non-mutator} is effectively beneficial.

Actual biological populations may not have all {the} mutations having the same selective effects. There are models in which {the} selection coefficient is chosen from a distribution. However, the robustness of the results presented here could be tested using other fitness functions. There have been works taking {into} account the possible physiological costs associated with lowering {the} mutation rates \citep{Kimura:1967,Kondrashov:1995,Dawson:1998,Johnson:1999a,Baer:2007}. The effect of this factor could be explored.

\subsection{Choice of parameters and biological relevance}\label{param_choice} 

\citet{Maisnier-Patin:2005} experimentally confirmed that in \textit{{Salmonella typhimurium}}, for various {values of the} mutation rates{, the} fitness effect of the mutations resembles the function (\ref{fitfunction}) with $\alpha$ $=$ $0.46$. Synergistic epistasis has been observed in experiments \citep{Mukai:1969,Whitlock:2000}. In \textit{{Drosophila melanogaster}}, the logarithm of relative productivity of genotypes was measured to be proportional to negative of the number of mutant regions carried by them \citep{Whitlock:2000}. This is similar to the fitness function (\ref{fitfunction}) with {the} corresponding $\alpha$ being $2$. In previous theoretical studies, the chosen values for $\alpha$ range from $0.02$ \citep{Fumagalli:2015} to $5$ \citep{Campos:2004}, whereas $\alpha$ has been varied from {$0.03$} to $20$ in the simulations in this article. The strength $\lambda$ of the mutator can be as large as $1000$ \citep{Miller:1996} to as small as around $2$ \citep{Mcdonald:2012,Wielgoss:2013}. In this study, $\lambda$ values ranging from $1.25$ to $10,000$ {(see Appendix} \ref{alphaC_wm}) are used. For \textit{{E. coli}} populations, $s$ is observed to be in the range $10^{-3}$ \citep{Gallet:2012} $-$ $10^{-1}$ \citep{Lenski:1991}, while $U_d$ $\sim$ $10^{-3}$ per genome per generation \citep{Drake:1998}. The present article includes values of $s${ and the} absolute values of mutation rate (for both mutators and {non-mutator}s) in the range $10^{-5}$ $-$ $10^{-1}$ and $10^{-4}$ $-$ $10^{-1}$ per genome per generation{, }respectively. In the experiment {by} \citet{Maisnier-Patin:2005}, $U_d$ was varied from $4$ $\times$ $10^{-4}$ to $0.31$ per genome per replication, and $s$ was measured to be $4.1$ $\times$ $10^{-2}$.

With the help of information about the parameters $s$, $U_d$ and $\alpha$ from the experiment of \citet{Maisnier-Patin:2005}, it follows from (\ref{mutfreq_0}) {that $p(0)$ lies between $0.99$ and} $7.3$ $\times$ $10^{-18}$ corresponding to the lower and upper limits of $U_d$. By including the selection coefficient, the condition proposed by \citet{Kondrashov:1993} can be modified to write $N_m p(0)s$ $\sim$ $10^2$, where $N_m$ is the minimum value of $N$ so as to have {a} steady state (see {Appendix} \ref{steadystate}). Hence, for the experiment of \citet{Maisnier-Patin:2005}, $N_m$ varies from $3$ $\times$ $10^3$ to $3$ $\times$ $10^{20}$. {Nevertheless}, the population size in this case was $10^8$. For this value of $N$, a steady state is possible for $U_d$ $\leq$ $0.16$. Therefore, to the best of our knowledge, \textit{{Salmonella typhimurium}} is the only model organism {which can be used to test} the results in this article, {as} it is the only asexual for which $\alpha$ has been measured.

\subsection{Fixation time and comparison with experiments}\label{exp_comp} 

The time required for the fixation of a {non-mutator} is given by the inverse of the rate at which {non-mutator}s that are certain to get fixed are created \citep{Weinreich:2005}. (The details of {the} fixation time can be found in \citet{Ewens:2004}.) The rate of creation of {non-mutator}s that are expected to reach fixation is the product of their rate of production and fixation probability. Thus, for a large population, {the} fixation time \citep{James:2016} {is} $T$ $=$ $(N b \Pi)^{-1}$, where $b$ is the rate at which the mutation that produces the lower mutation rate allele happens, provided $N b$ $\ll$ $1$. 

Table $2$ of \citet{Wielgoss:2013} gives the mutation rate corresponding to $3$ genotypes and their respective times of origin. Assuming the time of origin corresponds to the time when a genotype was significantly high in proportion in order to get detected, we see that the time for reduction of mutation rate is inversely proportional to magnitude of the reduction. {However}, the experiment of \citet{Mcdonald:2012} indicates that this reduction time is higher for {a} higher magnitude of decline in {the} mutation rate. The opposite trends observed in the above two experiments can be explained if epistasis is assumed to be present, as the fixation probability can be either {an} increasing or decreasing function of mutation rate{ depending} on {the} epistasis parameter. 

\subsection{{Comparison with previous theoretical works}}\label{theo_comp}

It is known that the fixation probability of an allele with effective selective advantage $S$ in a finite population of size $N$ is $\Pi(N)$ $=$ $(1-e^{-2S})/(1-e^{-2SN})$ \citep{Kimura:1962}. {In this model, for infinite population size,} $\Pi$ $=$ $2S$, as discussed in \citet{James:2016}. That is, the fixation probability of a beneficial allele in an infinite population is twice its net selective advantage. For a harmful allele, $S$ is negative, and hence, {its} fixation probability (time) falls (increases) exponentially with $N$ \citep{Kimura:1980,Assaf:2011}. 

Fixation of mutators in an asexual {non-mutator} population is effectively the same as fixation of a harmful allele if {the} fitness affecting beneficial mutations are excluded. For this problem, \citet{Jain:2012} studied the fixation time of mutators, and the time was found to {increase exponentially} ($e^{2N \sqrt{\frac{2 \alpha (\lambda-1)}{\pi} ~ s^{\frac{1}{\alpha}}~ \left(\frac{U_d}{\lambda}\right)^{2-\frac{1}{\alpha}}} }$ for weak selection, and ${e^{2N U_d ~ \left(1-1/\lambda\right)}}$ {for strong selection). It is important to be noted that we study only the strong mutator case ($\lambda$ $\gg$ $1$). From the above mentioned result of} \citet{Jain:2012}, {we can obtain the effective selective disadvantage conferred by the mutator (which is also the same as $(\Pi/2)$) to be} $\sqrt{\frac{2 \alpha \lambda}{\pi}} ~ s^{\frac{1}{2 \alpha}} ~ \left(\frac{U_d}{\lambda}\right)^{1-\frac{1}{2 \alpha}}$ {for weak selection.} This has similar dependence on $U_d$ as (\ref{fptot_exp_ws_al1}), though the mutator strength does not enter our expression. For $\alpha$ $=$ $1$, these two solutions differ only by a factor $2$. Nevertheless, {in the strong selection regime}, the net selective disadvantage of the mutator is simply {$U_d(1-1/\lambda)$}, which exactly matches with our result. In the case of the work of \citet{Jain:2012}, there is a continuous production of mutators from {non-mutator}s owing to {which} mutators sweep to fixation in a finite population. For a population of large size, the corresponding steady state fitness is $e^{-U_d}.$ In the present article, we analyze a mutator population that is {in steady state} initially with mean fitness $e^{-U_d}$. The {non-mutator} allele can appear in a background carrying $k$ mutations, and reach fixation to form a distribution of {non-mutator}s with mean fitness $(1-s)^{k^{\alpha}} e^{-U_d/\lambda}$. Though the initial state of the problem addressed in this article is the same as the final state of the problem considered by \citet{Jain:2012}, the reverse is not true. In the special case of strong selection, these two articles study ``complementary'' processes.

\section*{Acknowledgements}
The author is thankful to CSIR for the funding {as well as} K. Jain, K. Zeng, and B. Charlesworth for the discussions. The author is grateful to K. Jain for some valuable comments{ and} two anonymous reviewers for their suggestions. {The author extends his thanks to Vinutha L. for her help with proofreading and editing the manuscript.}

\clearpage


\begin{figure}
\begin{center} 
\includegraphics[width=1.0 \linewidth,angle=0]{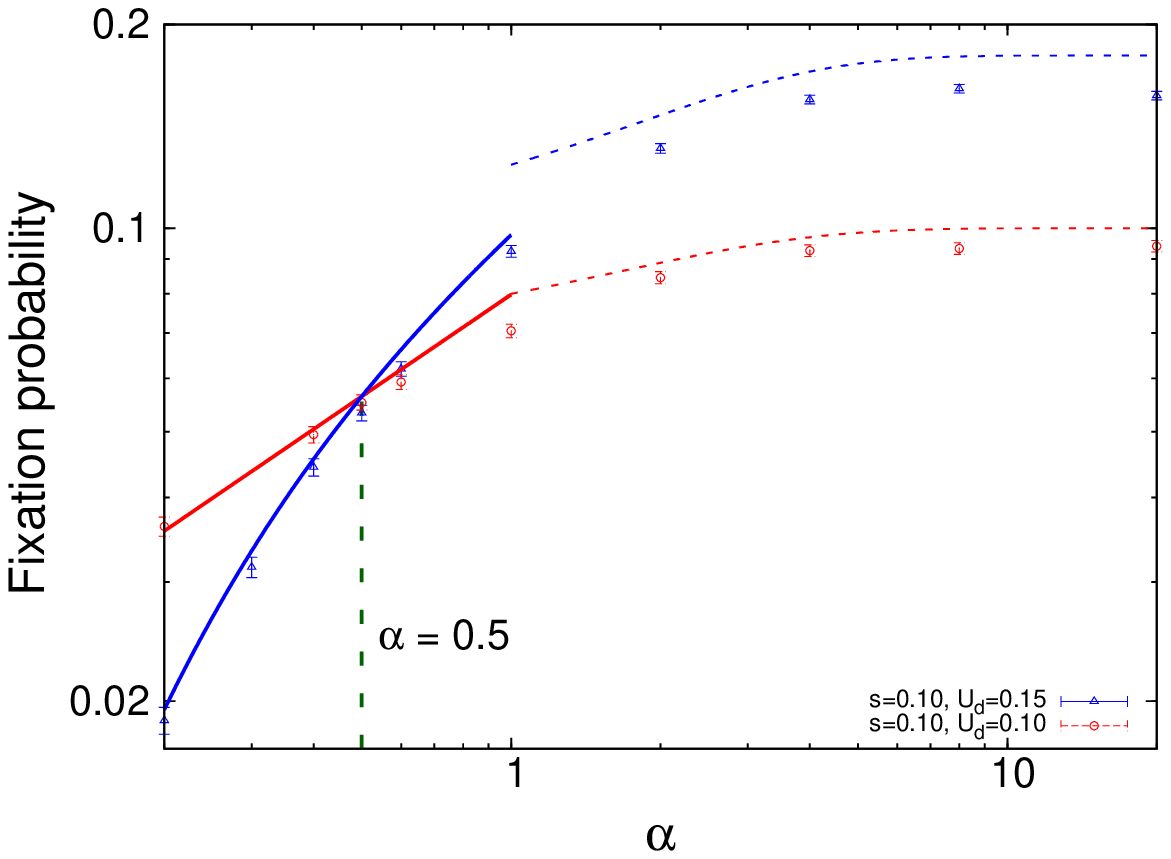}
\caption{Weak selection with antagonistic and synergistic epistasis. The symbols represent simulation data (red circles for $s$ $=$ $0.1$, $U_d$ $=$ $0.1${, and} $\lambda$ $=$ $100${; }blue triangles for $s$ $=$ $0.1$, $U_d$ $=$ $0.15${, and} $\lambda$ $=$ $100$). Each simulation point is averaged over $10^5$ independent stochastic runs. The error bars stand for {$\pm 2$ standard error} \citep{Cumming:2007}. The corresponding solid curves indicate (\ref{fptot_exp_ws_al1}), and the dashed curves represent (\ref{fptot_ws_ag1}). The green vertical broken line is drawn at $\alpha$ $=$ $0.5$.}  
\label{fpsm}
\end{center} 
\end{figure}

\begin{figure}
 \begin{center} 
 \includegraphics[width=1.0 \linewidth,angle=0]{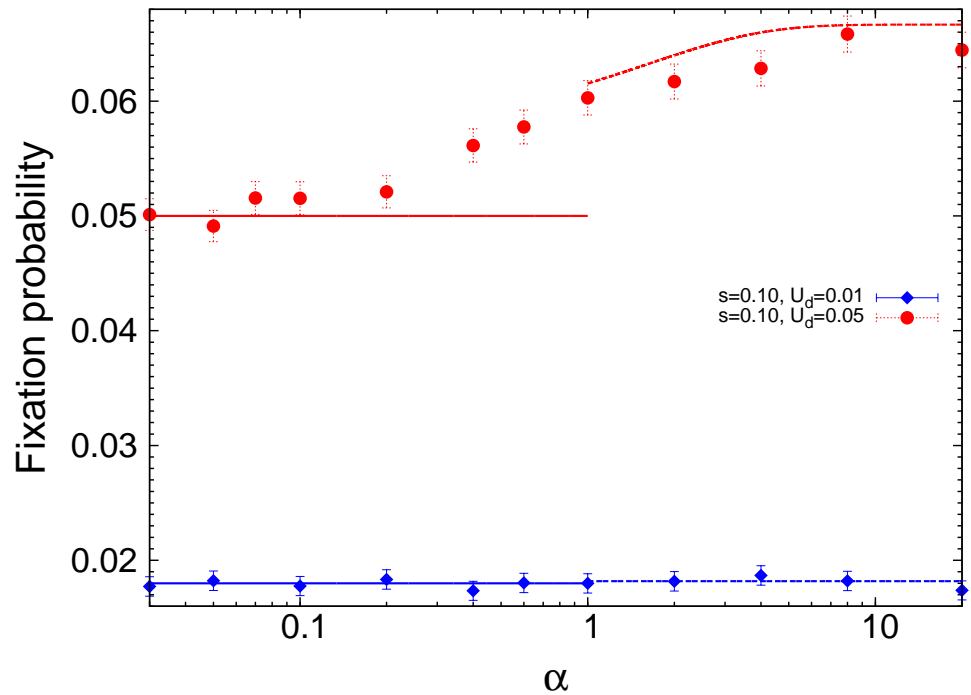}              
 \caption{Strong selection with antagonistic and synergistic epistasis. The symbols represent simulation data (red circles for $s$ $=$ $0.1${ and} $U_d$ $=$ $0.05${; } blue diamonds for $s$ $=$ $0.1${ and} $U_d$ $=$ $0.01$). $\lambda$ $=$ $100$ for both the cases. Each simulation point is averaged over $10^5$ independent stochastic realizations. The error bars represent {$\pm 2$ standard error}. The solid lines correspond to (\ref{fptot_ss_al1}), and the broken curves represent (\ref{fptot_ss_ag1}).}
\label{fpsm_ss}
\end{center}
\end{figure}

\begin{figure}
 \begin{center} 
 \includegraphics[width=1.0 \linewidth,angle=0]{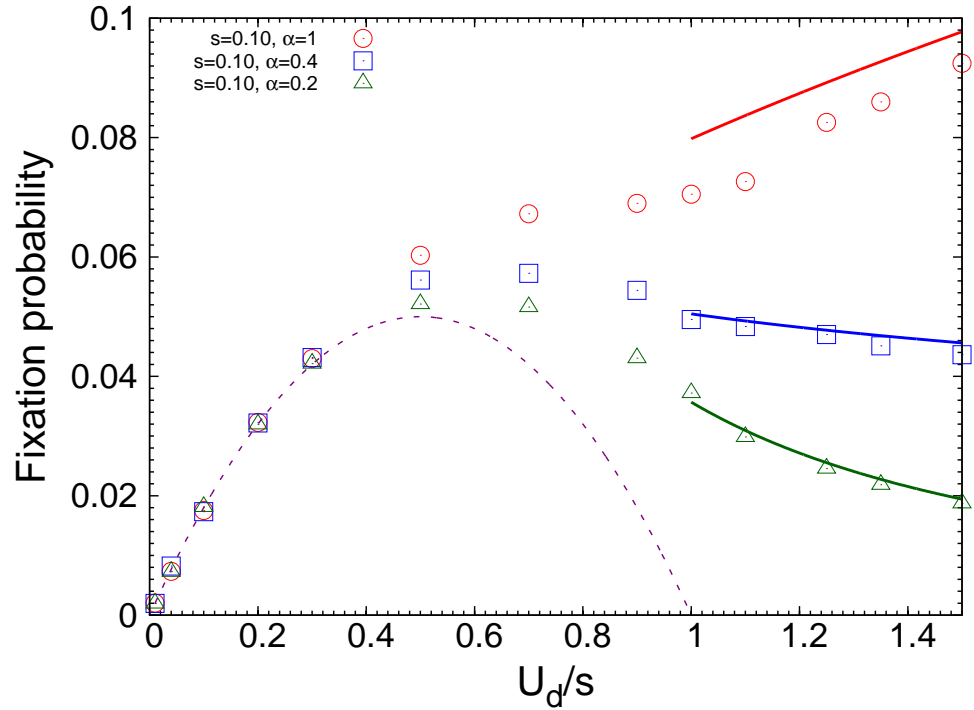}              
 \caption{Antagonistic epistasis with strong and weak selection. Variation of $\Pi$ with $U_d$. The symbols represent simulation data (red circles for $\alpha$ $=$ $1$, blue squares for $\alpha$ $=$ $0.4${, and} green triangles for $\alpha$ $=$ $0.2$). Each point is averaged over $10^5$ independent stochastic runs. The other parameters are $s$ $=$ $0.1$ and $\lambda$ $=$ $100$. The solid curves correspond to (\ref{fptot_exp_ws_al1}), and the broken curve represents (\ref{fptot_ss_al1}). Clearly, (\ref{fptot_ss_al1}) deviates from simulation results as $(U_d/s)$ $\rightarrow$ $1$.}
\label{fpnonmon_al1}
\end{center}
\end{figure}

\begin{figure}\begin{center} \includegraphics[width=1.0 \linewidth,angle=0]{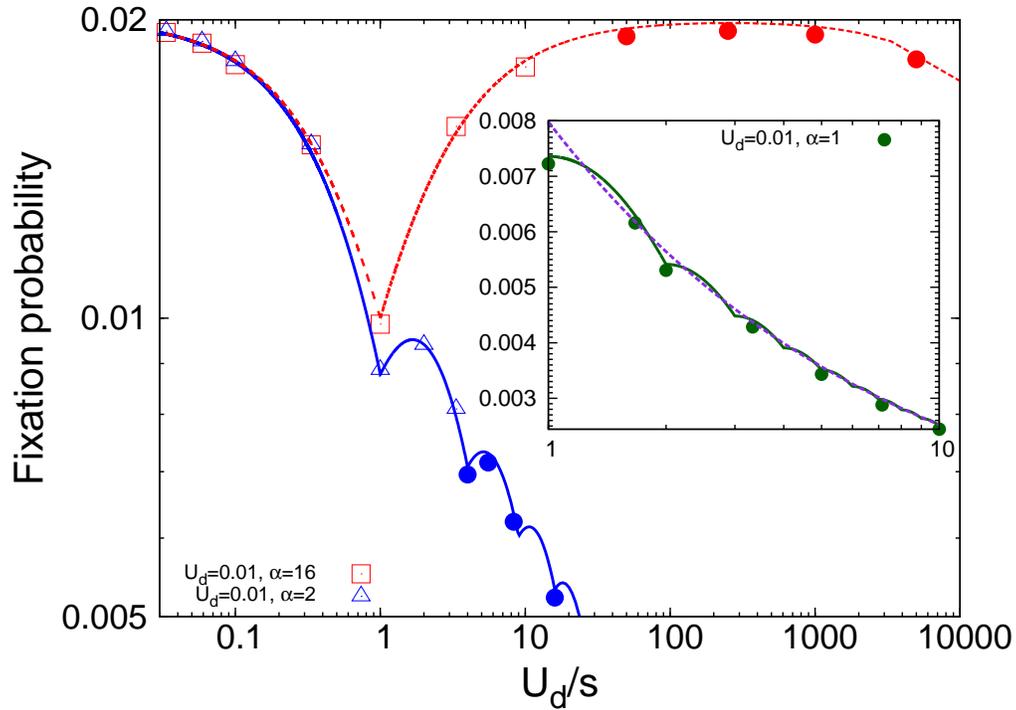}              
\caption{Synergistic epistasis with strong and weak selection. Variation of $\Pi$ with $s$. {The open symbols show simulation data} (averaged over $10^5$ independent realizations). The filled circles represent the numerical {solutions} of (\ref{fpk_rec}) and (\ref{fptot}). The parameters are $\alpha$ $=$ $16$ (red squares), $2$ (blue triangles){, and} $1$ (green circles) with $U_d$ $=$ $0.01$ and $\lambda$ $=$ $100$. The red curves plotted with the data respectively indicate (\ref{fptot_ws_ag1}) {and} (\ref{fptot_ss_ag1}) in {the weak and strong} selection regimes. The blue and green (inset) solid curves are (\ref{fp_a2}) and (\ref{fp_a1}){, respectively}. The broken violet line in the inset represents the simplified expression (\ref{fptot_exp_ws_al1}).}
\label{fps__ag1}\end{center}\end{figure}

\begin{figure}\begin{center} \includegraphics[width=1.0 \linewidth,angle=0]{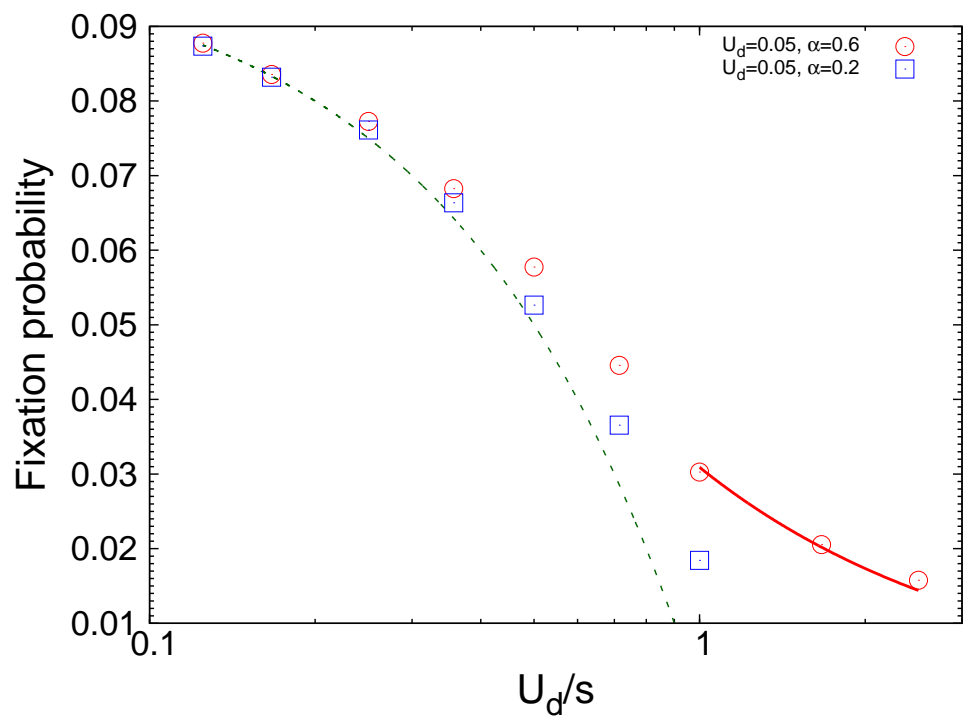}              
\caption{Antagonistic epistasis with weak and strong selection. Variation of $\Pi$ with $s$. {The symbols show simulation data} (average over $10^5$ replicas). {Here,} $\alpha$ $=$ $0.6$ (red circles) and $0.2$ (blue squares){. The other parameters are} $U_d$ $=$ $0.05$ and $\lambda$ $=$ $100$. The red solid curve is (\ref{fptot_exp_ws_al1}), {and the green broken curve is} (\ref{fptot_ss_al1}).}
\label{fps__al1}\end{center}\end{figure}

\clearpage


\appendix
\numberwithin{equation}{section}
\numberwithin{table}{section}
\numberwithin{figure}{section}

\section{Frequency of mutator population}
\label{mutfreq}

When $s$ and $U_d$ are small, {the population fraction of mutators carrying $k$ mutations in generation $t$ can be expressed using the equation} 
\be
\label{mutfreq_t}
{\frac{\partial p(k,t)}{\partial t} = U_d[\bar p(k-1,t)-\bar p(k,t)] - s[k^{\alpha} - \overline{k^{\alpha}}(t)]p(k,t)} ~,~
\ee
{where} 
\be
\label{meanmut}
{\overline{k^{\alpha}}(t) = \sum_{k=0}^{\infty} \left(k\right)^{\alpha} p(k,t)} ~.~
\ee
{When the population is in} steady state, (\ref{mutfreq_t}) and (\ref{meanmut}) become time independent, and we get
\be
\label{mutfreq_ss}
U_d[\bar p(k-1)-\bar p(k)] - s[k^{\alpha} - \overline{k^{\alpha}}]p(k) = 0 ~.~
\ee
Solving (\ref{mutfreq_ss}) for $k=0$ yields expression for negative of the mean Wrightian fitness (logarithm of {the} Malthusian fitness given by (\ref{fitfunction})) of the population per selection coefficient 
\be 
\label{meanmutexp}
\overline{k^{\alpha}} = U_d/s ~.~
\ee
{This} can be substituted back in (\ref{mutfreq_ss}){ and} iterated to get \citep{Jain:2008}
\be
\label{mutfreq_soln}
p(k) = \frac{\left(U_d/s\right)^{k}}{\left(k!\right)^{\alpha}} p(0) ~.~
\ee
The normalization condition $\sum_{k=0}^{\infty} p(k) = 1$ gives \citep{Jain:2008} 
\be
\label{mutfreq_0}
p(0) = \left[\sum_{k=0}^{\infty} \left(U_d/s\right)^{k} \left(k!\right)^{-\alpha} \right]^{-1} ~.~
\ee
{The exact} solutions to the above expression are possible {only} for $\alpha$ $=$ $1$ ({non-e}pistatic case) \citep{Kimura:1966,Haigh:1978} and $\alpha$ $=$ $2$ \citep{Jain:2008}.
\begin{equation}
{p(0) =
\begin{cases}
e^{-U_d/s} ~ ~ ~  \text{if} \hspace{0.05 cm} ~~ \alpha=1 \\
\left[I_0\left(2\sqrt{\frac{U_d}{s}}\right)\right]^{-1}~~ \text{if} \hspace{0.05 cm} ~~ \alpha=2.
\end{cases}}
\label{mutfreq_0_alpha}
\end{equation}
$I_0\left(2\sqrt{\frac{U_d}{s}}\right)$ is the modified Bessel function of the first kind of order $0$. (Refer to \citet{Abramowitz:1964} to know more about this function.) It is a monotonically increasing function of $(U_d/s)$, so that $p(0)$ decreases with increase in $(U_d/s)$. For any value of $\alpha$ except $1$ and $2$, approximations are needed to solve (\ref{mutfreq_0}).

\section{Approximate expressions for {the} class zero mutator frequency}
\label{mutfreq0}

By taking the ratio $p(k)/p(k-1)$ in (\ref{mutfreq_soln}), we can see that the maximum of $p(k)$ is at $k_m$ $=$ $(U_d/s)^{1/\alpha}$.

\underline{{Case I: $U_d/s$ $>$ $1$, $\alpha$ $\leq$ $1$}}

\begin{figure}
\begin{center} 
\includegraphics[width=1.0 \linewidth,angle=0]{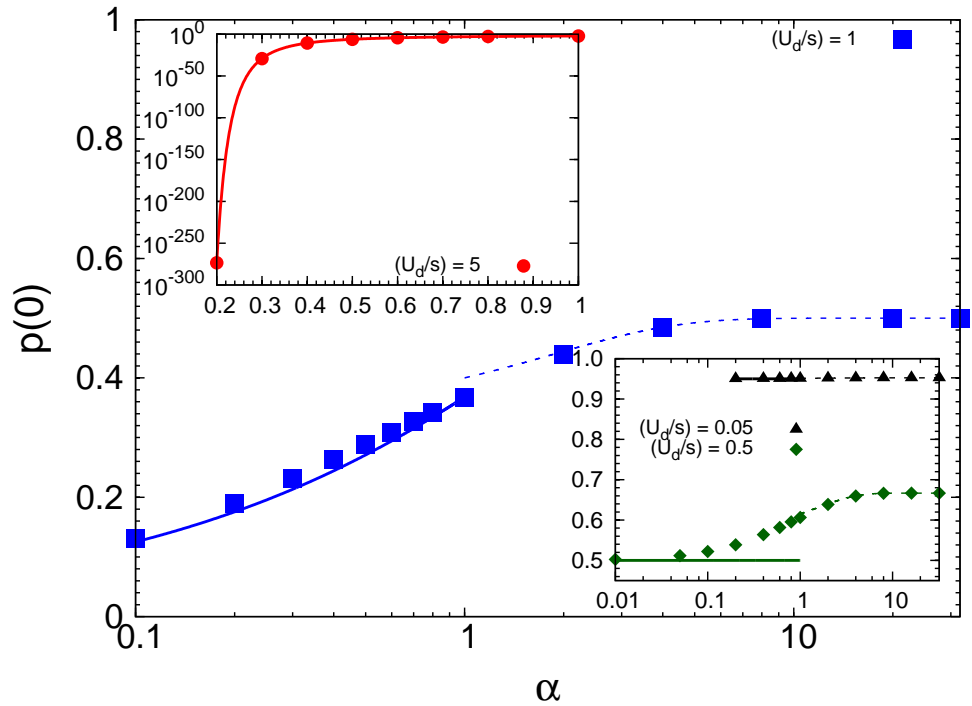}
\caption{{Class $0$ mutator fractions}. For the main figure and the insets, the symbols represent the numerically evaluated values of the full sum in (\ref{mutfreq_0}). Main figure: Weak selection with antagonistic and synergistic epistasis. Solid and broken curves are expressions (\ref{mutfreq_0_simple}) and (\ref{mutfreq_0_ag1}){, respectively} for $(U_d/s)$ $=$ $1$ (blue squares). Left top inset: Weak selection; antagonistic epistasis. The red solid curve is (\ref{mutfreq_0_simple}) for $(U_d/s)$ $=$ $5$. Right bottom inset: Strong selection with antagonistic and synergistic epistasis. The solid and broken curves{ show} (\ref{mutfreq_al1}) and (\ref{mutfreq_0_ag1_ss}){, respectively for $(U_d/s)$ $=$ $0.05$ (black triangles) and $(U_d/s)$ $=$ $0.5$ (green diamonds).}} 
\label{Q} 
\end{center} 
\end{figure}

{For $\alpha$ $\leq$ $1$, when selection is weaker than mutation rate,} $(U_d/s)^{1/\alpha}$ $>$ $1$. {Correspondingly,} the distribution of mutators can be approximated by a Gaussian. As a first step, upon using Stirling's approximation $k! \approx \sqrt{2\pi k} ~ (k/e)^k$ in (\ref{mutfreq_soln}), we obtain 
\be
\label{p0_stirling}
p(k) = \frac{\left(U_d/s\right)^{k} ~ e^{k\alpha}}{\left(\sqrt{2\pi k}~k^k\right)^{\alpha}} ~ p(0) ~.~
\ee
Now, converting (\ref{p0_stirling}) to an exponential{ and} then expanding around its maximum $(U_d/s)^{1/\alpha}$ using Taylor series gives 
\be
\label{p0_gaussian}
p(k) = \frac{e^{\alpha (U_d/s)^{1/\alpha}} }{\left(2\pi (U_d/s)^{1/\alpha}\right)^{\alpha/2} } ~ ~ e^{\frac{-\alpha \left(k-(U_d/s)^{1/\alpha}\right)^2}{2(U_d/s)^{1/\alpha}}} ~ p(0) ~.~
\ee
By replacing the sum in (\ref{mutfreq_0}) by an integral and using (\ref{p0_gaussian}), we get  
\be
\label{mutfreq_0_int}
p(0) = \left(\frac{ e^{\alpha (U_d/s)^{1/\alpha}}}{\left(\sqrt{2\pi (U_d/s)^{1/\alpha}}\right)^{\alpha}} \left[\int_{x=0}^{(U_d/s)^{1/\alpha}} e^{-\frac{\alpha x^2}{2 (U_d/s)^{1/\alpha}}} dx + \int_{x=0}^{\infty} e^{-\frac{\alpha x^2}{2 (U_d/s)^{1/\alpha}}} dx \right] \right)^{-1} .
\ee
Performing the integral in (\ref{mutfreq_0_int}) yields 
\be
\label{mutfreq_0_expr}
p(0) = \left(\frac{  e^{\alpha (U_d/s)^{1/\alpha}}}{\left(\sqrt{2\pi (U_d/s)^{1/\alpha}}\right)^{\alpha}  } \sqrt{\frac{\pi (U_d/s)^{1/\alpha}}{2 \alpha}} \left[ 1 + erf{\left(\sqrt{\frac{\alpha (U_d/s)^{1/\alpha}}{2}}\right)} \right] \right)^{-1} ~.~
\ee
For large values of $x$, we have the expansion $erf(x)$ $\approx$ $1-\frac{e^{-x^2}}{x \sqrt{\pi}}$ $\approx$ $1${. By using this,} we can simplify (\ref{mutfreq_0_expr}) to write 
\be
\label{mutfreq_0_simple}
p(0) = (2 \pi)^{\frac{\alpha-1}{2}} e^{-\alpha (U_d/s)^{1/\alpha}} {\alpha}^{1/2} (U_d/s)^{\frac{\alpha - 1}{2 \alpha}} ~.~ 
\ee
Note that (\ref{mutfreq_0_simple}) reproduces the known result $p(0)$ $=$ $e^{-U_d/s}$ for $\alpha$ $=$ $1$. Fig. \ref{Q} shows a comparison of (\ref{mutfreq_0_simple}) with (\ref{mutfreq_0}). The inset at the left top clearly indicates that (\ref{mutfreq_0_simple}) very well captures the exact sum even for {extremely} small values of $p(0)$.  

\underline{Case II: $U_d/s$ $>$ $1$, $\alpha$ $>$ $1$}

In this case, {the} Gaussian approximation does not hold good. {An approximate solution is possible for the limit case $(U_d/s)^{1/\alpha}$ $<$ $2$.} When $(U_d/s)^{1/\alpha}$ $<$ $2$, {the} mutator frequency peaks around $1$ (also, see {section} \ref{wsse}), and contributions to $p(k)$ from classes with $k$ $>$ $1$ are negligibly small. Nevertheless, in order to obtain a very accurate estimate of $p(0)$, terms up to second order can be retained. Effectively, we get 
\be
\label{mutfreq_0_ag1}
{p(0) \approx \frac{1}{1+U_d/s+\frac{(U_d/s)^2}{2^{\alpha}}} ~~ \text{if \hspace{0.05 cm}} ~~ \alpha > \frac{\ln{(U_d/s)}}{\ln{2}}} ~.~
\ee
As $\alpha$ increases, $p(0)$ rises to the constant value $(1+U_d/s)^{-1}$, which is the upper bound of {$p(0)$}. Fig. \ref{Q} shows that (\ref{mutfreq_0_ag1}) {matches well with} (\ref{mutfreq_0}) for large values of $\alpha$. For values of $\alpha$ that are not very large, other classes also contribute. As a result, $p(0)$ will be smaller than what is predicted by (\ref{mutfreq_0_ag1}).

\underline{{Case III: $U_d/s$ $<$ $1$, $\alpha$ $<$ $1$}}

{When $U_d/s$ $<$ $1$, $p(k)$ in} (\ref{mutfreq_soln}) {is a monotonically decreasing function of $k$.} In the limiting case $\alpha$ $\rightarrow$ $0$, (\ref{mutfreq_0}) becomes 
\be
\label{mutfreq_al1}
p(0) \approx \left[\sum_{k=0}^{\infty} \left(U_d/s\right)^{k} \right]^{-1} = (1-U_d/s), ~~ \text{if \hspace{0.05 cm}} ~~ \alpha \ll 1 ~.~
\ee
{However,} (\ref{mutfreq_al1}) only provides the lower bound of $p(0)$ in {the} presence of antagonistic epistasis when {the} selection is strong. When both $(U_d/s)$ and $\alpha$ are not very small compared to $1$, this expression is not accurate to obtain the exact values of $p(0)$ (see Table \ref{table_p0_ssal1}). The right bottom inset of Fig. \ref{Q} shows that $p(0)$ predicted by (\ref{mutfreq_0}) decreases to (\ref{mutfreq_al1}) for very small values of $\alpha$. It can be seen that $p(0)$ depends on $\alpha$ if the numerical value of $(U_d/s)$ is close to $1$. Expression (\ref{mutfreq_al1}) does not capture this dependence. When $(U_d/s)$ $\ll$ $1$, $p(0)$ becomes independent of $\alpha$. {Moreover}, Table \ref{table_p0_ssal1} and Fig. \ref{Q} indicate that corresponding to the same value of $\alpha${, as the value of $(U_d/s)$ increases, $p(0)$ decreases.}

Note that when $\alpha$ $=$ $0$, it follows from (\ref{fitfunction}) that all {the} individuals carrying {non-zero} mutations have the same fitness $(1-s)$. Thus, in practice, the population has only two classes differing in fitness by $s$, with mutation rate from class $0$ to $1$ being $U_d$. For this, the steady state solution for population fraction in class $0$ yields (\ref{mutfreq_al1}). 

\begin{table}
\begin{center}
  \begin{tabular}{| l || c | c | c | c | }
\hline
\multicolumn{4}{ |c| }{\bf{{Comparison of} (\ref{mutfreq_al1}) {with} (\ref{mutfreq_0})}} \\
    \hline
    {$\alpha$} & {$U_d/s$} \vspace{0.0 cm} & {$p(0)$ (exact)} & {$(1-U_d/s)$} \\ \hline \hline
  ${1}$  & {$0.9$} & {$0.407$} & {$0.1$}  \\ \hline
  ${10^{-1}}$  & {$0.9$} & {$0.196$} & {$0.1$}  \\ \hline
  ${10^{-2}}$  & {$0.9$} & {$0.115 $} & {$0.1$}  \\ \hline
  ${10^{-3}}$  & {$0.9$} & {$0.102 $} & {$0.1$}  \\ \hline \hline
  ${1}$  & {$0.5$} & {$0.607 $} & {$0.5$}  \\ \hline
  ${10^{-1}}$  & {$0.5$} & {$0.522 $} & {$0.5$}  \\ \hline
  ${10^{-2}}$  & {$0.5$} & {$0.503 $} & {$0.5$}  \\ \hline \hline
  ${1}$  & {$0.2$} & {$0.819 $} & {$0.8$}  \\ \hline
  ${10^{-1}}$  & {$0.2$} & {$0.803 $} & {$0.8$}  \\ \hline \hline
 ${1}$  & {$0.1$} & {$0.905$} & {$0.9$}  \\ \hline 

  \end{tabular}
\caption {Evaluation of $p(0)$ (exact) is done using (\ref{mutfreq_0}). For {small $(U_d/s)$ values}, (\ref{mutfreq_al1}) and (\ref{mutfreq_0}) show a good agreement. For larger $(U_d/s)$, these two solutions match {only} for $\alpha$ $\ll$ $1$.} \label{table_p0_ssal1}
\end{center}
\end{table}

\underline{Case IV: $U_d/s$ $<$ $1$, $\alpha$ $>$ $1$}

Like Case II studied here, {the} class {$0$} mutator fraction increases with $\alpha$.  
\be
\label{mutfreq_0_ag1_ss}
{p(0) \approx \frac{1}{1+U_d/s+\frac{(U_d/s)^2}{2^{\alpha}}} ~~ \text{if \hspace{0.05 cm}} ~~ \alpha > 1} ~.~
\ee
From the right bottom inset of Fig. \ref{Q}, we can see that $p(0)$ predicted by (\ref{mutfreq_0}) increases to {$(1+Ud/s)^{-1}$} for large values of $\alpha$. When $(U_d/s)$ $\ll$ $1$, (\ref{mutfreq_al1}) and (\ref{mutfreq_0_ag1_ss}) give almost the same result close to $1$ for $p(0)$, indicating the fact that {the} fraction of individuals with zero deleterious mutations in the population is unaffected by epistasis when {the} selective effects are very strong. Since the population is localized around class {$0$}, {the} frequency of individuals in other fitness classes will be insignificant. Note that when $(U_d/s)$ is not very small compared to $1$, $p(0)$ depends on $\alpha$. 

Results of this section are summarized in Table \ref{tablep0}.

\section{{Explanation for the trend in Fig.} \ref{fps__ag1}}
\label{trend_fig6}

{Table} \ref{table_synws} {gives the numerical values of $\pi(k)$ using} (\ref{fpk_rec}), {and $p(k)$ for different values of $s$. When $s$ is reduced, $\pi(0)$ remains roughly the same, while $p(0)$ decreases. This explains why $\Pi$ $=$ $\pi(0)p(0)$ falls as a function of $s$ in the strong selection regime. For $k$ $>$ $0$, for a given value of $k$,} $\pi(k)$ $=$ $2(U_d-sk^{\alpha})$ (see (\ref{fpk})) {increases with decrease in $s$. From Appendix} \ref{mutfreq0}, {we understand that the distribution $p(k)$ peaks around} $k_m$ $=$ $(U_d/s)^{1/\alpha}$. Column $2$ of Table \ref{table_synws} shows this value. The integer part $\lfloor k_m \rfloor$ of the corresponding number gives the maximum value of the fitness class that contributes to $\Pi$. In {the} weak selection regime, when $s$ is lowered keeping $\lfloor k_m \rfloor$ to be the same, we see from Table \ref{table_synws} that $p(k_m)$ remains almost the same, while $p(k)$ decreases for all $k$ $<$ $k_m$. Hence, the net effect on $\Pi$ due to decrease in $s$ is not a straightforward problem. To have a better understanding at least for $1$ $\leq$ $(U_d/s)^{1/\alpha}$ $<$ $2$ ($\lfloor k_m \rfloor$ $=$ $1$), we can make use of (\ref{fptot_ws_ag1}). {The argument used in understanding the case $\lfloor k_m \rfloor$ $=$ $1$ can be extrapolated to the general case $\lfloor k_m \rfloor$ $=$ $n$, where $n$ takes only positive integer values}. 

{By }taking the first derivative of $\Pi$ in (\ref{fptot_ws_ag1}) with regard to $s$, we can see that $\Pi$ peaks at a particular value of selection
\be
{s^{*} = 2^{-\alpha/2} U_d} ~.~
\label{sstar}
\ee 
Note that for $\alpha=2$, (\ref{sstar}) yields $s^*$ $=$ $5\times 10^{-3}$ ($U_d/s^*$ $=$ $2$), which approximately matches with the simulation data (blue triangles) shown in Fig. \ref{fps__ag1}, whereas for $\alpha=16$ (shown using {the} red squares), $s^*$ $=$ $3.9\times 10^{-5}$ ($U_d/s^*$ $=$ $256$). The latter point also agrees with the data plotted in Fig. \ref{fps__ag1}. The red curves for the {weak and strong selection regimes are} (\ref{fptot_ws_ag1}) {and} (\ref{fptot_ss_ag1}){, respectively}. $\Pi$ increases with decrease in $s$ for $U_d$ $<$ $s$ $\leq$ $s^*$. Further decrement in $s$ reduces $\Pi$ for $s^*$ $<$ $s$ $\leq$ $2^{-\alpha}U_d$. When $s$ is lowered from $n^{-\alpha}~U_d$ to $(n+1)^{-\alpha}~U_d$, where $n$ $=$ $1,2,3,...$ (see Table \ref{table_synws}), $\lfloor k_m \rfloor$ increases by $1$. Consequently, frequencies $p(k)$ of classes having high $\pi(k)$ values decrease. For $k$ $>$ $1$, $\pi(k)$ declines rapidly with $k$. This can be explicitly seen {in} the last row in Table \ref{table_synws}, for which {the} selection is very small. Therefore, each time $\lfloor k_m \rfloor$ increases by $1$ due to decrement in $s$, $\Pi$ increases initially, followed by a faster decay. {As a result}, whenever $(U_d/s)$ $=$ $n^{\alpha}$ with $n$ being any {non-zero} value of $\lfloor k_m \rfloor$, $\Pi$ assumes its local minimum values, as we see in Fig. \ref{fps__ag1}. 

In {the} weak selection regime, for $\alpha$ $>$ $1$, $\Pi$ {undergoes damped oscillations and decreases overall.} We will examine the properties of its local maxima or peaks. The relative increase in $\Pi$ corresponding to $s^*$ (first peak) can be measured as 
\be
{\frac{\Delta \Pi(s^{*})}{\Pi(s)|_{U_d}} = \frac{\Pi(s)|_{s^{*}}-\Pi(s)|_{U_d}}{\Pi(s)|_{U_d}}} ~.~
\label{pinon_sstar}
\ee
{Substituting} (\ref{sstar}) in (\ref{fptot_ws_ag1}), {we get} 
\be
{\Pi(s)|_{s^{*}} = \frac{2U_d~2^{\alpha/2}}{2+2^{\alpha/2}}} ~.~
\label{pi_sstar}
\ee
For $\alpha$ $=$ $16$, (\ref{pi_sstar}) yields $\Pi(s,\alpha=16)|_{s^{*}}$ $=$ $2U_d$, and (\ref{fptot_ws_ag1}) leads to $\Pi(s,\alpha=16)|_{U_d}$ $=$ $U_d$. Indeed, these two match with the observed values in Fig. \ref{fps__ag1}. Thus, we estimate the value of the relative increase in $\Pi$ corresponding to $s^*$ to be $100$ percent for $\alpha$ $=$ $16$ using (\ref{pinon_sstar}). This result is applicable for any large value of $\alpha$ for which $2^{\alpha/2}$ $\gg$ $2$. For $\alpha$ $=$ $2$, (\ref{sstar}) can be substituted back in the exact expression (\ref{fp_a2}) {to obtain} $\Pi(s,\alpha=2)|_{s^{*}}$ $=$ $0.94U_d$. At $s$ $=$ $U_d$, using (\ref{fp_a2}), $\Pi(s,\alpha=2)|_{U_d}$ {$=$ $0.88U_d$. These two results also are in good agreement with what is observed in Fig.} \ref{fps__ag1}. The resulting relative increase in $\Pi$ for $\alpha$ $=$ $2$ corresponding to $s^*$ is $6.8$ percent. It is evident that with reduction in $\alpha$, the first peak gets smaller. Moreover,  Fig. \ref{fps__ag1} suggests that for the same value of $\alpha$, the peaks associated with further reductions in $s$ ($\lfloor k_m \rfloor$ $>$ $1$) become less significant. 

For $\alpha$ $\leq$ $1$, the approximation $\lfloor k_m \rfloor$ $=$ $k_m$ was made in order to get {the} simple analytical expression (\ref{fptot_exp_ws_al1}). At least for $\alpha$ $=$ $1$, for which an exact formula for $p(0)$ is available, the presence of {non-m}onotonicity in {the} weak selection regime can be tested. Using (\ref{mutfreq_0_alpha}) in (\ref{fptot_full}), we can write 
\be
{\Pi = \frac{2U_d ~ \left(\frac{U_d}{s}\right)^{\lfloor (U_d/s) \rfloor}}{\left(\lfloor (U_d/s) \rfloor !\right)} ~ e^{-U_d/s}} ~.~
\label{fp_a1}
\ee
In the weak selection regime, when $\lfloor U_d/s \rfloor$ $=$ $n$, this can be rewritten as $\Pi$ $=$ $2U_d~(U_d/s)^n~ e^{-U_d/s}/(n!)$. By differentiating $\Pi$ with respect to $s$, one can see that $\Pi$ peaks at $s$ $=$ $n^{-1}U_d$. {However}, in this section, we {have already seen} that the local minima of $\Pi$ occur at $s$ $=$ $n^{-1}U_d$. This reflects the fact that there is no {non-m}onotonicity for $\alpha$ $=$ $1$. The inset of Fig. \ref{fps__ag1} shows (\ref{fp_a1}) using dark green lines, and the filled circles are the numerical solutions of $\Pi$ using (\ref{fpk_rec}) and (\ref{fptot}). The broken violet line is the approximate expression (\ref{fptot_exp_ws_al1}) for $\alpha$ $=$ $1$. For antagonistic epistasis, from the simulation data in Figures \ref{fpnonmon_al1} and \ref{fps__al1}, we do not see any {non-m}onotonic trend in {the} weak selection regime unlike Fig. \ref{fps__ag1}. Note that for the set of parameters used in these two figures, $\lfloor (U_d/s)^{1/\alpha} \rfloor$ can be as large as $7$.  

 \begin{table}
\resizebox{0.95\textwidth}{!}{\begin{minipage}{\textwidth}
\begin{center}
\begin{tabular}{ |l|l||l|l|l|l||l|l|l|l||l|l| }
\hline
\multicolumn{12}{ |c| }{\bf{{Synergistic epistasis: Variation of $\Pi$ with selection}}} \\
\hline
${\left(\frac{s}{U_d}\right)}$ & ${\left(\frac{U_d}{s}\right)^{\frac{1}{\alpha}}} $ & \multicolumn{4}{ |c|| }{${\pi(k)\times 10^{2}}$}  & {$p(0)$} & {$p(1)$} & {$p(2)$} & {$p(3)$} &  \multicolumn{2}{ c| }{${\Pi\times 10^{3}}$} \\ \hline
 & & {$k=0$} & {$k=1$} & {$k=2$} & {$k=3$} & & & & & {Exact} & {Eqn.} (\ref{fp_a2})\\ \hline \hline

 {$10.0$} & {$0.32$} & {$1.96$} & & & & {$0.91$} & & & & {$17.8 $} & {$18.1$} \\ \hline

 {$3.00$} & {$0.58$} & {$1.96$} & & & & {$0.73$} & & & & {$14.4 $} & {$14.7$} \\ \hline

 {$1.00$} & {$1.00$} & {$1.96$} & {$0.00$} & & & {$0.44$} & {$0.44$}& & & {$8.6 $} & {$8.8$} \\ \hline

 {$0.90$} & {$1.05$} & {$1.97$} &{$0.17$} & & & {$0.41$} & {$0.45$} & & & {$8.8 $} & {$9.0$} \\ \hline

 {$0.50$} & {$1.41$} & {$1.97$} &{$0.97$} & & & {$0.24$} & {$0.47$} & & & {$9.2 $} & {$9.4$} \\ \hline

 {$0.25$} & {$2.00$} & {$1.98$} &{$1.47$} &{$0.00$} & & {$0.09$} & {$0.35$} & {$0.35$} & & {$7.0 $} & {$7.1$} \\ \hline

 {$0.20$} & {$2.24$} & {$1.98$} &{$1.57$} & {$0.38$} & & {$0.06$} & {$0.29$} & {$0.37$} & & {$7.2 $} & {$7.3$} \\ \hline

 {$0.14$} & {$2.67$} & {$1.98$} &{$1.70$} & {$0.86$} & & {$0.03$} & {$0.19$} & {$0.34$} & & {$6.7 $} & {$6.9$} \\ \hline

 {$0.10$} & {$3.16$} & {$1.98$} &{$1.78$} & {$1.12$} & {$0.18$} & {$0.01$} & {$0.11$} & {$0.28$} &{$0.31$} & {$6.0 $} & {$6.1$} \\ \hline

\end{tabular}
 \caption {The data corresponds to $U_d$ $=$ $0.01$ and $\alpha$ $=$ $2$. The integer value corresponding to the number in the second column gives the number of classes that contribute to $\Pi$. $\pi(k)$ and $\Pi$ are scaled by $10^2$ and $10^3${, respectively}. $\pi(k)$ values are obtained numerically via solving (\ref{fpk_rec}), while $p(k)$ using (\ref{mutfreq_0_alpha}). To get the exact values of $\Pi$ given in column $11$, (\ref{fpk_rec}) and (\ref{fptot}) have been numerically solved.} 
\label{table_synws}
\end{center}
\end{minipage} }
\end{table}

\section{{Regarding the discrepancy between the analytical and simulation results}}
\label{limit_ana_res} 

To have {a} steady state, the size of a population needs to be of the order of $100 ~ (p(0))^{-1}$ \citep{Kondrashov:1993}. Table \ref{tablep1} gives $p(0)$ values by solving (\ref{mutfreq_0_simple}) corresponding to two $(U_d/s)$ values, changing $\alpha$. For large $(U_d/s)$ and small $\alpha$, we find that{ the} size required {for the attainment of} steady state is too large for most of the biological populations. Hence, populations of lower size will accumulate deleterious mutations{ and} go extinct (see section \ref{summary}). By comparing {columns} $4$ and $5$, it can be seen that as $\alpha$ decreases, (\ref{fptot_exp_ws_al1}) deviates from the exact solution of $\Pi$ obtained using (\ref{fpk_rec}) and (\ref{fptot}). This is because the approximation (\ref{fpk}) does not hold good for {the} fitness classes close to {$\lfloor$}$(U_d/s)^{1/\alpha}${$\rfloor$}, which contribute more to $\Pi$ due to the form (\ref{pkgaussain}) taken by {the} mutator frequency. Moreover, for a particular {value of $\alpha$, the deviation is less if the value of $(U_d/s)$ is small. Nevertheless,} populations of size in the biological limit having larger values of $(U_d/s)$ and smaller {values of} $\alpha$ do not have steady state. {Hence}, (\ref{fptot_exp_ws_al1}) is applicable to most of the real populations except those with both $(U_d/s)$ $\sim$ $1$ and antagonistic epistasis with very small $\alpha$ values. However, a better approximation to $\pi(k)$ is needed to yield more accurate results for $\Pi$ when $(U_d/s)$ $\gg$ $1$ and $\alpha$ $\ll$ $1$. 

One more thing to note is that $\alpha_c$ $=$ $0.5$ is obtained using (\ref{fptot_exp_ws_al1}). This tells us that the ``true'' value of $\alpha_c$ could be slightly different from $0.5$, as (\ref{fptot_exp_ws_al1}) deviates from the exact results. The {best estimate} of $\alpha_c$ is given in {Appendix} \ref{alphaC_wm}. 

\begin{table}
\begin{center}
  \begin{tabular}{| c || c | c | c | c | }
\toprule
\multicolumn{5}{ |c| }{\bf{{Weak selection; antagonistic epistasis:}}} \\
\multicolumn{5}{ |c| }{\bf{{Comparison of} (\ref{fptot_exp_ws_al1}) {with the exact numerical solution for $\Pi$}}} \\
    \hline
    $\alpha$ & $U_d/s$ \vspace{0.0 cm} & $p(0)$ & {Exact value of} $\Pi \text{\hspace{0.00 cm}}$ & $\Pi \text{\hspace{0.00 cm}}$ {using} (\ref{fptot_exp_ws_al1}) \\  
  ~ & ~ & ~ & using (\ref{fpk_rec}) and (\ref{fptot}) & ~ \\ \hline \hline
  $0.5$  & $5$ & $7.44 \times10^{-7}$ & ${1.03 \times10^{-2}}$ &  ${1.19 \times10^{-2}}$ \\ \hline
  $0.45$  & $5$ & $1.56\times10^{-8}$ & ${8.00 \times10^{-3}}$ &  ${8.96 \times10^{-3}}$ \\ \hline
  $0.4$  & $5$ & $2.12\times10^{-11}$ & ${5.84 \times10^{-3}}$ &  ${6.75 \times10^{-3}}$ \\ \hline
  $0.3$  & $5$ & $6.24\times10^{-30}$ & ${2.13 \times10^{-3}}$ &  ${2.99 \times10^{-3}}$ \\  \hline \hline
  $0.2$  & $1.5$ & ${2.09 \times10^{-2}}$ & ${3.81 \times10^{-3}}$ &  ${3.89 \times10^{-3}}$ \\  \hline
  $0.15$  & $1.5$ & ${5.99 \times10^{-3}}$ & ${2.23 \times10^{-3}}$ &  ${2.40 \times10^{-3}}$ \\  \hline
  $0.1$  & $1.5$ & $6.98\times10^{-5}$ & ${7.62 \times10^{-4}}$ &  ${9.97 \times10^{-4}}$ \\  
\bottomrule  
\end{tabular}
\caption {{Here,} $p(0)$ is evaluated using (\ref{mutfreq_0_simple}). The value of $s$ is chosen to be $0.02$ for the two values of $(U_d/s)$ used.} \label{tablep1}
\end{center}
\end{table} 

To derive (\ref{fpk}), we assumed that $\pi(k)$ $\ll$ $1${ and} neglected cubic and higher order terms. When $U_d$ $\gtrsim$ $0.1$, $\pi(k)$ values will differ from what we obtain using (\ref{fpk}). Due to the same reason, there is discrepancy between (\ref{fptot_ws_ag1}) and (\ref{fptot_ss_ag1}), {as well as} the simulation data in Figures \ref{fpsm} and \ref{fpsm_ss}{, respectively}. The same expressions match well with {the} simulation points for $U_d$ $=$ $10^{-2}$ as shown in Fig. \ref{fps__ag1}{ and} using the blue diamonds in Fig. \ref{fpsm_ss}. Corresponding to the point $\alpha$ $=$ $20$, Table \ref{table_pi0} provides a comparison of (\ref{fptot_ws_ag1}) and (\ref{fptot_ss_ag1}), and the simulation results in Figures \ref{fpsm} and \ref{fpsm_ss}{, respectively}. In fact, (\ref{fptot_exp_ws_al1}) also deviates from the simulation results for large $U_d$ values. This is clear in the case of the red circles corresponding to $\alpha$ $=$ $1$ and the blue squares corresponding to $\alpha$ $=$ $0.4$ in Fig. \ref{fpnonmon_al1}, and in Fig. \ref{fpsm}, where $log$ scale is used to plot the data. {Moreover}, based on the discussion in {Appendix} \ref{trend_fig6}, it is worth noting that (\ref{fptot_full}) is the more accurate form of (\ref{fptot_exp_ws_al1}), but the latter is more simplified.

\begin{table}
\begin{center}
  \begin{tabular}{| l || l | l | l | }
    \hline
\multicolumn{4}{ |c| }{\bf{{Synergistic epistasis: Comparison of simulation data in }}} \\
\multicolumn{4}{ |c| }{\bf{{Figures} \ref{fpsm} and \ref{fpsm_ss} {with analytical results}}} \\
    \hline
 {$U_d$} & $\overline{\Pi}_{sim}$ {(standard error)} & ${\Pi_{num}}$ & ${\Pi_{analytic}}$ \\ \hline \hline

 {$0.15$} &	${1.571\times 10^{-1} ~ (1.15\times 10^{-3})}$ & ${1.553\times 10^{-1}}$ & ${1.80\times 10^{-1}}$ \\ \hline 
 {$0.10$} &	${9.406\times 10^{-2} ~ (9.231\times 10^{-4})}$ & ${9.12\times 10^{-2}}$ & ${1.00\times 10^{-1}}$ \\ \hline\hline
 {$0.05$} &	${6.445\times 10^{-2} ~ (7.765\times 10^{-4})}$ & ${6.33\times 10^{-2}}$ & ${6.67\times 10^{-2}}$ \\ \hline
 {$0.01$} & 	${1.738\times 10^{-2} ~ (4.133\times 10^{-4})}$ & ${1.785\times 10^{-2}}$ & ${1.82\times 10^{-2}}$\\ \hline
 \end{tabular}
\caption {The parameters used here are $\alpha$ $=$ $20$, $s=0.1${, and} $N$ $=$ $4000$. Note that $\Pi_{analytic}$ is given by (\ref{fptot_ws_ag1}) for weak selection, which is applicable to the $U_d$ values in the first two rows. $\Pi_{analytic}$ is obtained using (\ref{fptot_ss_ag1}) for strong selection, applicable to the $U_d$ values in the last two rows. $\Pi_{num}$ is obtained by solving (\ref{fpk_rec}) and (\ref{fptot}).} 
\label{table_pi0}
\end{center}
\end{table} 

The analytical expression (\ref{fptot_ss_al1}) for {the} strong selection; antagonistic epistasis regime (section \ref{ssae}) is valid only for $\alpha$ $\ll$ $1$. A more precise formula for $p(0)$ in this case can help us {to} analytically understand the variation of $\Pi$ with $\alpha$, and the {non-m}onotonic behavior in Fig. \ref{fpnonmon_al1}. {Further,} (\ref{fptot_ws_ag1}) and (\ref{fptot_ss_ag1}) are not exact expressions. A more accurate expression for $p(0)$ in {the} presence of synergistic epistasis will pave the way for better analytical understanding.

\section{{Regarding steady state}}
\label{steadystate} 

The argument of \citet{Kondrashov:1993} on the minimum population size necessary to ensure steady state includes only the population fraction corresponding to the least loaded class {(see section} \ref{sto_sim}). A further detailed analysis by \citet{Jain:2008} shows that the ratchet time is actually proportional to the number of individuals in the least loaded class times the selection coefficient. A very high ratchet time corresponds to very a slowly operating ratchet. For smaller values of selection coefficient, the deviation from {the results of} \citet{Kondrashov:1993} becomes clearer. Nevertheless, for the parameters used in this article, there will not be any significant difference from the above theory, {since} smaller values of selection have been used only for synergistic epistasis. 

\section{{Critical value of epistasis for the weak mutator background}}
\label{alphaC_wm} 

\begin{figure}\begin{center} \includegraphics[width=1.0 \linewidth,angle=0]{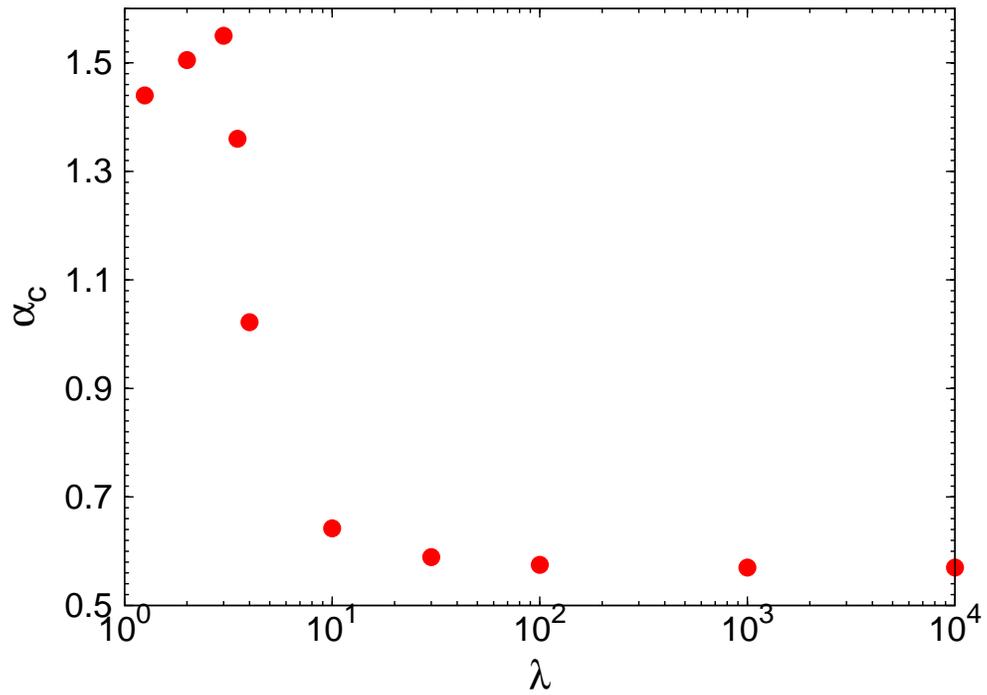}
\caption{{Variation of $\alpha_c$ with the strength $\lambda$ of the mutator. The points (red filled circles) are obtained using numerical solution of} (\ref{fpk_rec}) and (\ref{fptot}). {For all these points, $s$ $=$ $0.1$.}}  
\label{crit_alpha}\end{center} \end{figure}

Weak mutator refers to the case when the mutation rate of the {non-mutator} is comparable ($\lambda$ $\sim$ $1$) with that of the mutator. In two recent mutation reduction experiments \citep{Mcdonald:2012,Wielgoss:2013}, weak mutators with $\lambda$ as low as $2$ have been observed. By solving (\ref{fpk_rec}) and (\ref{fptot}) using \textit{{Wolfram Mathematica}} $9.0.1.0$, we obtain $\Pi$ as a function of $\alpha$ for a given value of $\lambda$. By comparing these $\Pi$ values up to {three} significant figures for two different values of $U_d$, we estimate $\alpha_c$ corresponding to each $\lambda$. With {the} variation in mutator strength, $\alpha_c$ changes. This is plotted in Fig. \ref{crit_alpha}. For strong mutators, the exact numerical analysis suggests that $\alpha_c$ approaches its minimum value $0.57$, as $\lambda$ $\rightarrow$ $\infty$. This is contrary to the analytical result $\alpha_c$ $=$ $0.5$ using (\ref{fptot_exp_ws_al1}). Therefore, even though (\ref{fptot_exp_ws_al1}) is helpful in demonstrating the presence of $\alpha_c$ (see section \ref{wsae}), this expression is not very accurate in determining $\alpha_c$ precisely. This is because of the two approximations (\ref{mutfreq_0_simple}) and (\ref{fpk}) involved in the derivation of (\ref{fptot_exp_ws_al1}) {(see Appendix} \ref{limit_ana_res} as well). As $\lambda$ falls towards $3$, $\alpha_c$ rises to its upper limit $1.55$. A further reduction in $\lambda$ results in decrease in $\alpha_c$. 

The interpretation for the initial increase of $\alpha_c$ with decrease in $\lambda$ is as {follows:} A significantly high deleterious mutation rate reduces the fixation probability of {non-mutator}, since it is a disadvantageous factor. Thus, a {non-mutator} produced in {a} weak mutator background that is less spread out, and that created in {a} strong mutator background which is more spread out can have the same fixation probabilities. The former and latter respectively correspond to larger and smaller values of $\alpha$. Therefore, the critical value $\alpha_c$ of {the} epistasis parameter rises as the strength of the mutator decreases. {However}, the decline of $\alpha_c$ with $\lambda$ for $\lambda$ $<$ $3$ is counterintuitive. The explanation for this interesting trend requires a detailed analysis. A study of the weak mutator case is beyond the scope of this article{ and} will be left for a separate work. The results presented regarding weak mutators is meant to give directions for future work. Solving (\ref{fpk_rec}) without neglecting $U'_d$ will be useful in obtaining analytical expressions in this case.

Table \ref{tableweakmut} gives the data from finite $N$ simulations for {three} values of $\lambda$. Using this, a crude estimate of $\alpha_c$ has been made corresponding to each $\lambda$. The errors associated with $\alpha_c$ are not given, since the error calculation is not straightforward here. Nevertheless, we see a clear variation with respect to $\lambda$, similar to what we see from the exact numerical solution.

\begin{table}
\begin{center}
\begin{tabular}{ |l|l|l|l||l|l|l|l| }
\hline
\multicolumn{8}{ |c| }{\bf{{Simulation results for weak mutators}}} \\
\hline
{$\lambda$} & {$U_d$} & {$\alpha$} & ${\overline{\Pi}_{sim} ~ (SE\times 10^{4})}$ & ${\overline{\Pi}_{sim} ~ (SE\times 10^{4})}$ & {$\alpha$} & {$U_d$} & {$\lambda$} \\ \hline
 &\multirow{7}{*} & {$1.7$} & ${0.04164 ~ (6.317)}$ & ${0.04159 ~ (6.314)}$ &  {$1.7$} & &  \\
{$2$} & {$0.15$} & \textbf{{1.6}} & ${0.04023 ~ (6.214)}$ & ${0.04025 ~ (6.215)}$ & \textbf{{1.6}} & {$0.10$} & {$2$} \\ 
 & & {$1.5$} & ${0.03866 ~ (6.096)}$ & ${0.04048 ~ (6.232)}$ & {$1.5$} & & \\ \hline

 &\multirow{7}{*} & {$1.0$} & ${0.05411 ~ (7.154)}$ & ${0.05253 ~ (7.055)}$ &  {$1.0$} & &  \\
{$4$} & {$0.15$} & \textbf{{0.9}} & ${0.05140 ~ (6.983)}$ & ${0.05106 ~ (6.961)}$ & \textbf{{0.9}} & {$0.10$} & {$4$} \\ 
 & & {$0.8$} & ${0.04659 ~ (6.665)}$ & ${0.04931 ~ (6.847)}$ & {$0.8$} & & \\ \hline

 &\multirow{7}{*} & {$0.6$} & ${0.06195 ~ (7.623)}$ & ${0.05926 ~ (7.466)}$ &  {$0.6$} & &  \\
{$100$} & {$0.15$} & \textbf{{0.5}} & ${0.05332 ~ (7.105)}$ & ${0.05527 ~ (7.226)}$ & \textbf{{0.5}} & {$0.10$} & {$100$} \\
 & & {$0.4$} & ${0.04432 ~ (6.508)}$ & ${0.04953 ~ (6.861)}$ & {$0.4$} & & \\ \hline
\end{tabular}
 \end{center}
\caption {{Example of values for $\Pi$ for two different values of $U_d$, and $\alpha_c$ is obtained as the value of $\alpha$ for which the difference between the two $\Pi$ values is the minimum.} The $\alpha_c$ value corresponding to each $\lambda$ is indicated in bold. The simulation parameters are $s$ $=$ $0.1$, $N$ $=$ $4000$, and each simulation point is averaged over $10^5$ independent stochastic realizations. The standard error ($SE$) value given in each row is the actual value multiplied by $10^4$. The {best estimates of $\alpha_c$ using} \textit{{Mathematica}} (see Appendix \ref{alphaC_wm} and Fig. \ref{crit_alpha}) are $1.505$, $1.022$ and $0.575$ for $\lambda$ $=$ $2$, $4$ and $100${, respectively}.} 
\label{tableweakmut}
\end{table}

\section{{Supplementary figures}}
\label{sup_fig}

\begin{figure}
\begin{center} 
\includegraphics[width=1.0 \linewidth,angle=0]{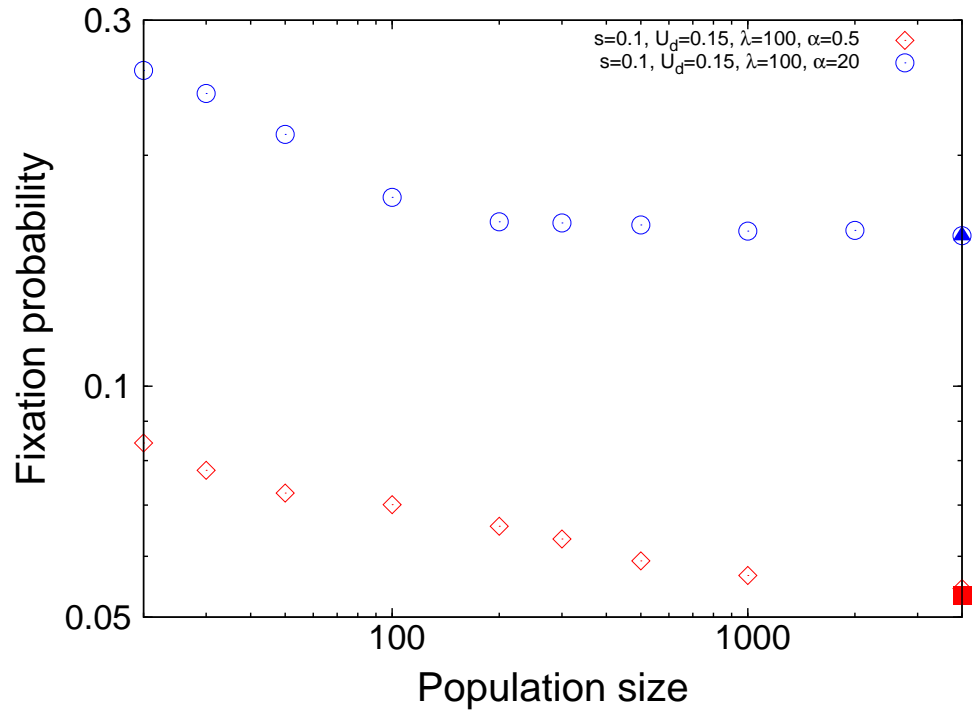}
\caption{Variation of the fixation probability with population size. A {non-mutator} is allowed to appear at time $t=10/s$ in a population that is initially at its fittest genotype. For large {values of} $N$ ($=$ $4,000$), {the} result in this case (shown using the open symbols) is in good agreement with the fixation probability of a {non-mutator} arising in a population which is initially {in} steady state (filled symbols). The parameters are $s$ $=$ $0.1$, $U_d$ $=$ $0.15$, $\lambda$ $=$ $100$ with $\alpha$ $=$ $20$ (blue {symbols}), and $\alpha$ $=$ $0.5$ (red {symbols}).}  
\label{fpNsm}
\end{center} 
\end{figure}

\begin{figure}\begin{center} \includegraphics[width=1.0 \linewidth,angle=0]{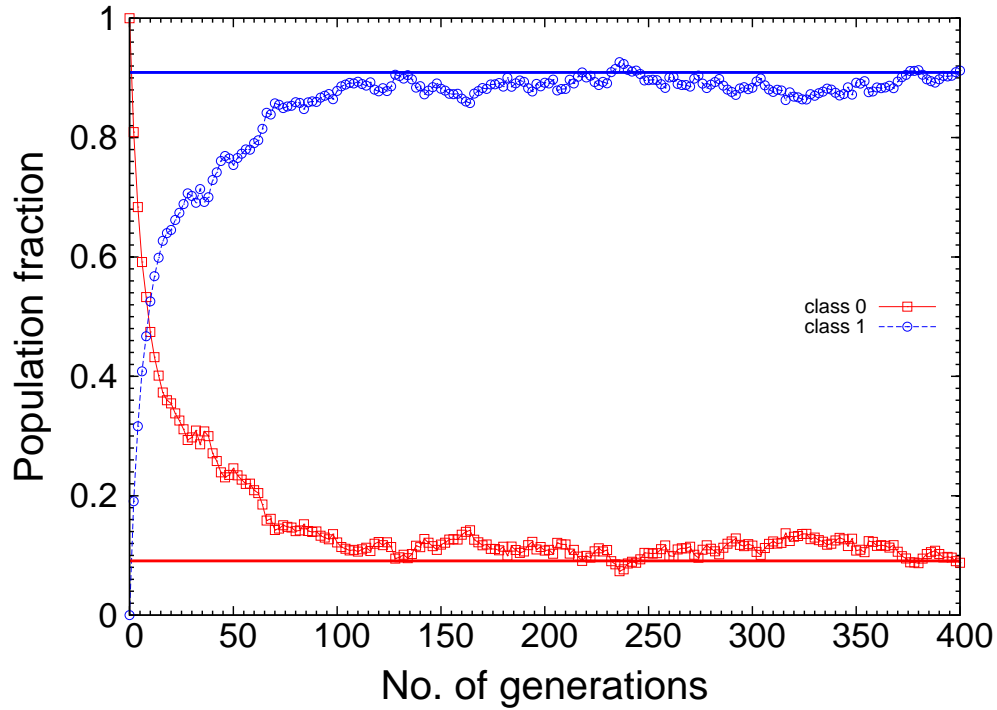}              
\caption{Weak selection; synergistic epistasis. Single run plot of population fractions. {It can be seen that the population will be localized in the first two fitness classes.} The parameters are $\alpha$ $=$ $16$, $U_d$ $=$ $0.1$, $s$ $=$ $0.01${, and $N$ $=$ $4,000$}. The solid horizontal lines show {the} steady state population fractions as calculated using (\ref{mutfreq_0}). At $t$ $=$ $0$, the population has no deleterious mutations. One can see that the population fractions from simulation approach the steady state values as time increases.}
\label{syn_cl}\end{center}\end{figure}

\clearpage



\end{document}